\documentclass[journal,10pt]{IEEEtran}

\usepackage[utf8]{inputenc}
\usepackage[T1]{fontenc}
\usepackage{cite}             
\usepackage{amsmath}
\usepackage{mathtools}
\usepackage{amssymb}
\usepackage{xifthen}
\usepackage{comment}

\usepackage{graphicx}
\usepackage[caption=false,font=footnotesize]{subfig}

\usepackage{algorithmicx}       
\usepackage{algorithm}          
\usepackage{algpseudocode}

\usepackage{amsthm} 
\newtheorem{theorem}{\bfseries Theorem}
\newtheorem{lemma}{\bfseries Lemma}
\newtheorem{definition}{\bfseries Definition}
\newtheorem{corollary}{\bfseries Corollary}

\newtheorem{remark}{\bfseries Remark}

\newtheorem{example}{\bfseries Example}

\newtheorem*{assumptions*}{\bfseries Assumptions}

\DeclareMathOperator{\diag}{diag}
\DeclareMathOperator{\eig}{eig}
\DeclareMathOperator{\IID}{i.i.d.}

\DeclareMathOperator*{\argmin}{arg\,min}

\newcommand{\realR}{\mathbb{R}}

\newcommand{\p}[1]{p_{X}}
\newcommand{\q}[1]{h^{#1}}
\newcommand{\Q}[1]{Q_{Y|X}^{#1}}
\newcommand{\Qm}[1]{q_{Y}^{#1}}
\newcommand{\va}[1]{
    \ifthenelse{\isempty{#1}} 
        {v} 
        {v^{#1}}
}
\newcommand{\tq}[1]{\hat{h}^{#1}}
\newcommand{\tQ}[1]{\hat{Q}_{Y|X}^{#1}}
\newcommand{\tQm}[1]{\hat{q}_{Y}^{#1}}
\newcommand{\tW}{\hat{W}}
\newcommand{\df}{\partial f}

\newcommand{\setSym}{\mathcal{X}}
\newcommand{\setY}{\mathcal{Y}}
\newcommand{\setX}{\mathcal{X}}

\newcommand{\A}[2][]{
    \ifthenelse{\isempty{#1}} 
        {A^{#2}(x,y,s)} 
        {A^{#2}(x,#1,s)}
}

\newcommand{\tA}[2][]{
    \ifthenelse{\isempty{#1}} 
        {\hat{A}^{#2}(x,y,s)} 
        {\hat{A}^{#2}(x,#1,s)}
}

\newcommand{\bA}[2][]{
    \ifthenelse{\isempty{#1}} 
        {A^{#2}[u](x,y,s)} 
        {A^{#2}[u](x,#1,s)}
}

\newcommand{\dsum}[2]{ 
    \sum_{(#1,#2) \in \setSym^2}
}

\newcommand{\ssum}[1]{ 
    \sum_{#1 \in \setSym}
}

\newcommand{\ssumh}[1]{ 
    \sum_{#1 \in \setSym}
}

\newcommand{\E}[2][]{ 
    \mathbb{E}_{#1} \left[ #2 \right]
}

\newcommand{\cx}[1][]{
    \ifthenelse{\isempty{#1}} 
        {c(y)} 
        {c(#1)}
}

\allowdisplaybreaks

\begin{document}

\title{Alternating Minimization Schemes for Computing Rate-Distortion-Perception Functions with $f$-Divergence Perception Constraints} 

\author{
   \IEEEauthorblockN{
        Giuseppe Serra, \IEEEmembership{Graduate Student Member, IEEE},
        Photios A. Stavrou, \IEEEmembership{Senior Member, IEEE}, 
        and Marios Kountouris, \IEEEmembership{Fellow, IEEE}}
  \thanks{Part of this paper was presented in the IEEE International Symposium on Information Theory (ISIT), Taiwan, 2023 \cite{serra:2023}.}
    \thanks{The authors are with the Communication Systems Department at EURECOM, Sophia-Antipolis, France, email: \{\texttt{giuseppe.serra, fotios.stavrou, marios.kountouris\}@eurecom.fr}. M. Kountouris is also with the Andalusian Research Institute in Data Science and Computational Intelligence (DaSCI), Department of Computer Science and Artificial Intelligence, University of Granada, Spain. }
    \thanks{The work of G. Serra and M. Kountouris is supported by the European Research Council (ERC) under the European Union’s Horizon 2020 Research and Innovation Programme (Grant agreement No. 101003431). The work of P. A. Stavrou is supported in part by the SNS JU project 6G-GOALS under the EU’s Horizon programme (Grant Agreement No. 101139232) and by the Huawei France-EURECOM Chair on Future Wireless Networks.}
  }


\maketitle

\begin{abstract}
 We study the computation of the rate-distortion-perception function (RDPF) for discrete memoryless sources subject to a single-letter average distortion constraint and a perception constraint belonging to the family of $f$-divergences. In this setting, the RDPF forms a convex programming problem for which we characterize optimal parametric solutions. We employ the developed solutions in an alternating minimization scheme, namely Optimal Alternating Minimization (OAM), for which we provide convergence guarantees. Nevertheless, the OAM scheme does not lead to a direct implementation of a generalized Blahut-Arimoto (BA) type of algorithm due to implicit equations in the iteration's structure. To overcome this difficulty, we propose two alternative minimization approaches whose applicability depends on the smoothness of the used perception metric: a Newton-based Alternating Minimization (NAM) scheme, relying on Newton's root-finding method for the approximation of the optimal solution of the iteration, and a Relaxed Alternating Minimization (RAM) scheme, based on relaxing the OAM iterates. We show, by deriving necessary and sufficient conditions, that both schemes guarantee convergence to a globally optimal solution. We also provide sufficient conditions on the distortion and perception constraints, which guarantee that the proposed algorithms converge exponentially fast in the number of iteration steps. We corroborate our theoretical results with numerical simulations and establish connections with existing results. 
\end{abstract}

\section{Introduction}
The rate-distortion-perception (RDP) trade-off studies the relevant problem of lossy compression under perceptual constraints on the reconstructed samples, generalizing the classical rate-distortion (RD) analysis \cite{shannon:59}. Concurrently proposed by Blau and Michaeli in \cite{blau:2019} and Matsumoto in \cite{matsumoto:2018, matsumoto:2019}, its introduction is motivated by the increasing necessity of a theoretical framework able to incorporate observations raised by a wide body of research spanning from machine learning to multimedia applications; see e.g.,  \cite{theis:16:hyperRes, shaham:18, kudo:19, minnen:18}, highlighting the presence of an inherent tension between perceptual quality and fidelity of the compressed samples. In this context, perceptual quality refers to the property of a sample to appear pleasing from a human standpoint. Empirical evaluations of perceptual quality using human scoring \cite{moorthy:2011, agustsson:2019:generative, mentzer:2020} show that conventional fidelity measures fail to capture human preferences and perceptions, especially in extreme compression scenarios. Hence, classical rate-distortion theory, focusing solely on the fidelity of the reconstructed samples, may not provide an adequate theoretical framework when applied to ``human-oriented'' data sources, such as images, audio, and video. 
\par The mathematical representation of the RDP trade-off is embodied by the rate-distortion-perception function (RDPF), which complements the classical single-letter rate-distortion function (RDF) with a divergence constraint between the source and reconstruction distributions. The additional constraint acts as a proxy for human perception, measuring the deviation from the real source distribution, also referred to as ``natural scene statistic'', following similar principles to those behind a class of no-reference image quality metrics \cite{mittal:2013, saad:2012}. However, it is worth noting that the selection of specific divergence metrics may be application-dependent and is still an active area of research. 
\par An alternative interpretation of the divergence constraints can be found in its potential as a semantic quality metric, i.e., a quantification of the importance of the reconstructed source to the observer \cite{kountouris:20}.  
For example, in \cite{katakol:2021}, a comparison between the segmentation capabilities of models trained on traditionally compressed samples against compressed samples with enhanced perceptual quality through Generative Adversarial Network (GAN)-based restoration shows a remarkable improvement in segmentation performance, especially for smaller scene objects.

\subsection{Related Work}\label{subsec:literature_review}
Since its introduction, the RDP trade-off has received substantial interest from the information theory community, deriving operational characterizations in a variety of operational scenarios. Focusing on the case where infinite common randomness is available both at the encoder and the decoder, Theis and Wagner \cite{theis:2021} provide variable-length codes for both the one-shot and asymptotic regime, exploiting the properties of the strong functional representation lemma \cite{li:2018}. 
In the context of the output-constrained RDF, but also valid for the "perfect realism" RDPF case, Saldi {\it et al.} \cite{saldi:2015} provide coding theorems for the case where only finite common randomness between encoder and decoder is available.
In \cite{chen:2022}, Chen {\it et al.} derive coding theorems for the asymptotic case, focusing on the differences between three operational cases defined by the availability of randomness between encoder and decoder, i.e., infinite common randomness, only private randomness, and the deterministic case.

\par Similar to its classical counterpart, the RDPF does not enjoy any general analytical solution. However, for specific source distributions and particular choices of distortion and perception measures, there exist closed-form expressions, such as for binary sources subject to Hamming distortion and total variation (TV) distance \cite{blau:2019} and Gaussian sources under mean squared-error distortion and various perception measures \cite{serra:2024, zhang:2021, qian:2025}. The available closed forms, while providing theoretical insights into the RDP trade-off, have limited applicability when considering arbitrary distortion and perception metrics. In the case of the classical RDF, the introduction of the celebrated Blahut-Arimoto (BA) algorithm \cite{blahut:72} mitigates the absence of a general closed-form solution, allowing for the exact computation of the RDF for the discrete case. Furthermore, its versatility inspired various approximation methods for the RDF in the continuous case \cite{yang:2021, yang:2024} and enabled its adaptation to a variety of source coding problems, with applications from quantum information theory \cite{Ramakrishnan:2020} to goal-oriented communication \cite{stavrou:2023:journal}. Moreover, in the RDPF case, numerical solutions have been explored to some extent. Data-driven solutions have been proposed, usually employing a generative adversarial scheme minimizing a linear combination of distortion and perception metrics, see, e.g., \cite{blau:2019, zhang:2021, kirmemis:2021, salehkalaibar:2024}. Despite the advantage of directly optimizing the image/video codec using only source samples, these methodologies still require considerable effort, as they are generally highly computationally demanding and data-intensive and may suffer from a lack of generalization capabilities. 
Algorithmic solutions for estimating the RDPF also exist. In the case of discrete alphabets, Chen {\it et al.} in \cite{chen:2023} cast the RDPF problem as an entropic-regularized Wasserstein barycenter problem and propose a solution method based on the Sinkhorn algorithm applicable to arbitrary distortion measures and with the perception measure being either a Wasserstein-type distance, the Kullback–Leibler divergence, or the TV distance. Focusing on Gaussian sources, Serra \textit{et al.} in \cite{serra:2024} design an alternating minimization method for the computation of the Gaussian RDPF for arbitrary fidelity and perception measures, deriving analytical solutions for the so-called ``perfect realism'' case. Furthermore, for the ``perfect realism'' case, Serra \textit{et al.} \cite{serra:2024:copula} design a solution algorithm for general multivariate continuous sources and distortion metrics, leveraging the information-geometric aspects of the constrained rate-distortion problem. 
To the best of our knowledge, none of the existing methodologies address the generic computation of the RDPF for discrete sources relying on generalizations of the classical BA algorithm.

\subsection{Contributions}
The objective of this work is to propose a generic algorithmic approach for the computation of the RDPF, focusing on the case of discrete memoryless sources subject to a single-letter average distortion constraint and a perception constraint belonging to the class of $f$-divergences.

Our results leverage the fact that the RDPF forms a convex program under mild regularity conditions on the perception constraint (specifically, convexity in the second argument), which are satisfied by the considered class of divergences. This enables us to derive a parametric characterization of the optimal solution of the RDPF (Lemma \ref{lemma:DoubleMinimization}), which is subsequently utilized to construct an alternating minimization procedure,
hereafter referred to as the Optimal Alternating Minimization (OAM) scheme, for which we also establish convergence guarantees (Theorem \ref{th:OAM}). 
However, the resulting structure of the OAM scheme relies on a set of implicit equations in the variables of interest, thus preventing the direct implementation of a generic BA algorithm, as is already known for the classical rate-distortion theory for $\IID$ sources and single-letter distortions \cite{blahut:72}.
Motivated by this technical difficulty, we propose two alternative minimization approaches that address the implementability issue and whose applicability depends on the smoothness of the considered perception function. 
\begin{itemize}
    \item In Section \ref{sec: NAM}, we design a \textit{Newton-based Alternating Minimization} (NAM) scheme observing that the solution of the OAM iterate is equivalent to a root-finding problem (Lemma \ref{lemma: Newton-Function Definition}), which allows us to apply Newton's root-finding method \cite{burden:2015} to compute the optimal iteration step (Theorem \ref{th: Newton Approximation Method}).
    \item In Section \ref{sec: RAM}, we introduce a \textit{Relaxed Alternating Minimization} (RAM) scheme, where we leverage a new relaxed formulation of the structure of the OAM iterations and subsequently, we derive necessary and sufficient conditions to ensure convergence to a globally optimal solution (Theorem \ref{th:RAM}). 
\end{itemize}
In Section \ref{sec: Algorithms Analysis}, we design the algorithmic implementations of the NAM and RAM schemes (see Algorithm \ref{alg: Newton Approximation} and Algorithm \ref{alg: RAM}, respectively) and develop suitable stopping criteria for both algorithms (Theorem \ref{th: UpperAndLowerboundToRate}). Moreover, we also provide sufficient conditions on the structure of the distortion and perception constraints under which our algorithms converge exponentially fast in the number of iterations (Theorems \ref{th: EXPConvergenceBase} and
\ref{th: approximation_exponential_interval}). We corroborate our theoretical findings with numerical simulations (Section \ref{sec: Numerical Examples}), with emphasis on the TV perception metric, for which we develop a smooth approximation (Lemma \ref{lemma: LowboundTV}).

\subsection{Notation} 
Let $\mathbb{N}$ denote the set of natural numbers, $[a:b] \subset \mathbb{N}$ the integer interval including its extremes, $\realR$ the set of real numbers, and $\realR_0^+$ the set of non-negative real numbers.
Given a discrete alphabet $\setSym$, we denote by $(\setSym, \mathbb{B}(\setSym))$ the Borel measurable space induced by the metric, with $\mathcal{P}(\setSym)$ denoting the set of probability measures defined thereon. We indicate with $\mathcal{Q}(\setSym)$ the set of all transition matrices $Q$ such that $Q \cdot p \in \mathcal{P}(\setSym^2)$ for all $p \in \mathcal{P}(\setSym)$.   We denote by $\E[]{\cdot}$ the expectation operator, and by $\E[q]{\cdot}$ we specify the probability distribution $q$ on which the expectation operator is applied. We indicate with square brackets the functional dependency between mathematical objects, e.g. $p[h]$ and $Q[h]$ express the functional dependence of a distribution $p \in \mathcal{P}(\setSym)$ or a transition matrix $Q \in \mathcal{Q}(\setSym)$ on another distribution $h \in \mathcal{P}(\setSym)$. We denote with $C^{n}$ the set of $n^{th}$-times differentiable functions. Given a function $f \in C^0$, we denote with $\df$ its sub-gradient \cite[Definition 8.3]{rockafellar:1998}, while, if $f \in C^2$, we denote by $f''(\cdot)$ the second derivative with respect to its argument. We denote by $D(\cdot||\cdot)$ a generic divergence measure, whereas $D_{f}(\cdot||\cdot)$ denotes a divergence belonging to the class of $f$-divergences.
Given a vector $v \in \mathbb{R}^d$, we indicate with $\diag{v}\in \mathbb{R}^{d \times d}$ the matrix with as diagonal the elements of $v$ and zeros otherwise. Given a matrix $V \in \mathbb{R}^{d \times d}$, we denote with $\eig(V)$ its set of eigenvalues.

\section{Preliminaries}
We start this section by providing the formal definition of the RDPF and an overview of its operational meaning, following \cite{blau:2019, theis:2021}. Subsequently, given their relevance to this work, we introduce the category of statistical divergences known as $f$-divergences and conclude with an overview of the alternating minimization methodology.

\subsection{Rate-Distortion-Perception Functions}
We consider finite alphabet sources and stochastic encoder/decoder pairs having access to a common source of randomness and define the minimum achievable rates under per-letter expected distortion and per-letter perception constraints. 

We assume that we are given an $\IID$ sequence of $n$-length random variables ${X}^n \in \setSym^n$ that induce the probability distribution $\p{} \in \mathcal{P}(\setSym)$. 
Formally, a stochastic encoder $f^n_E$ is any function in the set $\mathcal{F}^n_E = \{ f : \setSym^n \times \realR \to \mathbb{N} \}$, whereas a stochastic decoder $g^n_D$ is any function in the set $\mathcal{G}^n_D = \{ g: \mathbb{N} \times \realR \to \setSym^n \}$. A stochastic code is an element of $\mathcal{F}^n_E \times \mathcal{G}^n_D$. Without loss of generality, the randomness at the encoder and decoder is modeled as a single real number (i.e., representing an infinite number of bits), and is assumed shared by the pair, i.e., common randomness. 

We let $d: \setSym^2 \to \realR_0^+$ denote a single-letter distortion function and $D: \mathcal{P}(\setSym)^2 \to \realR_0^+$ denote a divergence function.
Moreover, we define the sets of fidelity criteria $\{\Delta_i\}_{i \in [1:n]}$ and $\{\Phi_i\}_{i \in [1:n]}$ as follows
\begin{align}
\Delta_i \triangleq \E[p_{X_i,Y_i}]{ d(X_i,Y_i)}\nonumber,  \qquad
\Phi_i \triangleq D( p_{X_i} || q_{Y_i})\nonumber
\end{align}
where $\Delta_i$ is the expected distortion of the $i^{th}$ symbol and $\Phi_i$  is the $i^{th}$ symbol divergence with respect to the reconstructed symbol $Y_i$. We are now ready to introduce the definition of achievability and that of the infimum of all achievable rates.

\begin{definition}(Achievability)
Given a distortion level $D \ge 0$ and a perception constraint $P \ge 0$, a rate $R$ is said to be $(D,P)$-achievable if there exists a random variable $U$ and a sequence of codes $(f_E^n,g_D^n) \in \mathcal{F}^n_E \times \mathcal{G}^n_D$ with
\begin{align*}
    K_n = f^n_E(X^n, U), \quad Y^n = g_D^n(K_n, U)
\end{align*}
such that, for $i = 1,\ldots,n$, the joint distribution $p_{X_i, Y_i}$ satisfies $\Delta_i \le D$ and $\Phi_i \le P$ and 
\begin{align*}
    \lim_{n \to \infty} \frac{H(K_n|U)}{n} \le R.
\end{align*}
Then, we define
\begin{align*}
    R_{cr}(D,P) \triangleq \inf\{R: R \text{ is $(D,P)$-achievable}\}.
\end{align*}
\end{definition}
Next, we state the definition of the information-theoretic characterization of the RDPF \cite{blau:2019}. 
\begin{definition}(RDPF)\label{def:rdpf}
    For a given finite alphabet source $X$ with distribution $\p{} \in \mathcal{P}(\setSym)$, a single-letter distortion $d(\cdot,\cdot)$ and a divergence $D(\cdot||\cdot)$, the RDPF is characterized as follows
    \begin{align}
        R(D,P) =        & \min_{\Q{} \in \mathcal{Q}(\setSym)}  I(X,Y)              \label{opt: primal}\\        \textrm{s.t.}   & \quad \E{d(X,Y)} \le D \label{opt:RDPF:DistortionConstraint}\\
                        & \quad D(\p{}||\Qm{}) \le P \label{opt:RDPF:PerceptionConstraint} 
    \end{align}
where $D\in[D_{\min},D_{\max}]\subseteq [0,\infty)$, $P\in[P_{\min},P_{\max}]\subseteq [0,\infty)$, $\Qm{} = \ssum{x} \Q{} \p{}$, and
\begin{align*}
     I(X,Y) = D_{KL}(\p{} \Q{}||\p{} \Qm{}) \triangleq I(\p{}, \Q{}) 
\end{align*}
where $I(\p{}, \Q{})$ highlights the dependency on $\{\p{},\Q{}\}$.
\end{definition}
In what follows, we highlight certain functional properties of Definition \ref{def:rdpf}.
\begin{remark}\label{remark:1}(On Definition \ref{def:rdpf} -  Functional properties in $(D,P)$) Following \cite{blau:2019}, it can be shown that \eqref{opt: primal} has some useful properties, under mild regularity conditions.
In particular, \cite[Theorem 1]{blau:2019} showed that, for $D\in[D_{\min},D_{\max}]\subset[0,\infty)$ and $P\in[P_{\min},P_{\max}]\subset[0,\infty)$, $R(D,P)$ is (i) monotonically non-increasing in both $D$ and $P$; (ii) convex in both $D$ and $P$ if the divergence $D(\cdot||\cdot)$ is convex in its second argument. 
\end{remark}
\begin{remark} \label{remark:2}  (On Definition \ref{def:rdpf} - Functional properties in $\Q{}$) The program defined by \eqref{opt: primal}-\eqref{opt:RDPF:PerceptionConstraint} is convex in the transition matrix $\Q{}$ for a given $p_X$ if the divergence $D(\cdot||\cdot)$ is convex in its second argument, since \eqref{opt: primal} and \eqref{opt:RDPF:DistortionConstraint} are respectively convex and affine functions in $\Q{}$ \cite{Csiszar1974}. Furthermore, the identity kernel $\Q{} = Id$ always satisfies the constraints given by \eqref{opt:RDPF:DistortionConstraint}-\eqref{opt:RDPF:PerceptionConstraint}.
\end{remark}

\par In the sequel, we assume that in \eqref{opt: primal}, { the perception constraint is an $f$-divergence}, i.e., $D(\cdot||\cdot)=D_f(\cdot||\cdot)$, which is known to be convex in both arguments \cite[Lemma 4.1]{fdiv-csizar}. 
\par We conclude this section by providing a theorem that connects $R_{cr}(D,P)$ with $R(D,P)$ for general alphabets.
\begin{theorem} For $D \ge 0$, $P \ge 0$, we obtain $R_{cr}(D,P)=R(D,P)$.
\end{theorem}
\begin{IEEEproof}   
 See \cite[Theorem 3]{theis:2021}.
\end{IEEEproof}

\subsection{Statistical divergences and the family of f-divergences.} \label{sec: f-divergences}
Statistical divergences are fundamental measures used in information theory and statistics to quantify the dissimilarity between probability distributions. In their general definition, a divergence on $\mathcal{P}(\setSym)$ is a function $D: \mathcal{P}(\setSym)^2 \to \realR_{0}^+$ such that $D(p||q) \ge 0$ for all $p,q \in \mathcal{P}(\setSym)$, holding with equality if and only if $p = q$.
Given the scope of this work, we focus on the family of $f$-divergences, first introduced in \cite{renyi:1961} (see also \cite{fdiv-csizar}). This rich class of divergences is defined as follows.

\begin{definition}{($f$-divergence)}
    Let $f:(0,\infty) \to \realR$ be a convex function with $f(1) = 0$. Then the $f$-divergence $D_f(\cdot||\cdot)$ associated with $f$ is defined as
    \begin{align*}
        D_f(p||q) \triangleq \ssum{x} q(x)f\left(\frac{p(x)}{q(x)}\right), \qquad  p,q \in \mathcal{P}(\setSym)
    \end{align*}
    under the assumption that
    \begin{align*}
        \text{(i) } f(0) &= \lim_{ x \to 0^+} f(0),  \qquad \text{(ii) } 0f\left( \frac{0}{0} \right) = 0, 
        \\ &\text{(iii) } \forall a \ge 0,~0f\left( \frac{a}{0} \right) = af'(\infty).  
    \end{align*}

\end{definition}
Many commonly used divergence functions belong to the class of $f$-divergences. For example,
\begin{itemize}
    \item KL divergence $D_{KL}(\cdot||\cdot)$, obtained by considering $f(x) = x\log(x)$,
        \begin{align*}
            D_{KL}(p||q) = \ssum{x} p(x)\log\left(\frac{p(x)}{q(x)}\right)
        \end{align*}
    \item Jensen-Shannon divergence $D_{JS}(\cdot||\cdot)$, where $f = x\log \left(\frac{2x}{x+1} \right) + \log \left(\frac{2}{x+1} \right)$,
        \begin{align*}
            D_{JS}(p||q) =  D_{KL}\left(p \Big|\Big|\frac{p+q}{2}\right) + D_{KL}\left(q\Big|\Big|\frac{p+q}{2}\right)
        \end{align*}
    \item  $TV(\cdot||\cdot)$, where $f = \frac{1}{2}|x - 1|$,
        \begin{align*}
            TV(P||Q) =  \frac{1}{2} \ssum{x} |p(x) - q(x)|
        \end{align*}
     \item $\alpha$-divergence $D_{\alpha}(\cdot||\cdot)$, where $f_{\alpha}$ is parameterized by $\alpha \in \mathbb{R}$,
        \begin{align*}
            D_{\alpha}(p||q) &= \ssum{x} q(x)f_{\alpha}\left(\frac{p(x)}{q(x)}\right) \qquad
            \\f_{\alpha}(x) &= \
            \begin{cases}
                \frac{x^\alpha - \alpha x - (1-\alpha)}{\alpha(\alpha -1)} \qquad \text{if } \alpha \neq 0, \alpha \neq 1 \\ 
                x\log(x) - x + 1 \qquad \text{if } \alpha = 1\\
                -\ln(x) + x - 1 \qquad \text{if } \alpha = 0\\
            \end{cases}.
        \end{align*}
\end{itemize}

We now state some general properties of this family of divergences. For any $f$-divergence $D_f(\cdot||\cdot)$, the following properties hold:
\begin{itemize}
    \item \textit{(Linearity)} $D_{f_1 + f_2}(\cdot||\cdot) = D_{f_1}(\cdot||\cdot) + D_{f_2}(\cdot||\cdot) $
    \item \textit{(Joint Convexity)} for any $t \in [0,1]$ and $p_1,p_2, q_1,q_2 \in \mathcal{P}$,
    \begin{align*}
        &D_f(tp_1 + (1-t)p_2||tq_1 + (1-t)q_2)  \\ 
        &\qquad \qquad\le tD_f(p_1||q_1) + (1-t)D_f(p_2||q_2).
    \end{align*}
    \item \textit{(Invariance)} Let $\hat{f}(x) = f(x) + c(x-1)$ for $c \in \realR$, then $D_f(\cdot||\cdot) = D_{\hat{f}}(\cdot||\cdot)$.
\end{itemize}

The characterization of the family of $f$-divergences provided here summarizes the properties useful for the scope of this work. For a more in-depth mathematical analysis, we refer the reader to \cite{sason:2018}.

\subsection{Alternating Minimization and BA-type algorithms} \label{sec: AM and BA-type algorithms}
The alternating minimization method is a framework for the minimization of functions of two constrained variables. Consider the following optimization problem
\begin{align*}
    \min_{\substack{ x \in \setX, y \in \setY }}  f(x,y)
\end{align*}
where $\setX$ and $\setY$ are two arbitrary non-empty sets and the function $f(x,y)$ satisfies $-\infty < f(x,y) \le +\infty$ for each $x\in\setX$ and $y\in\setY$. Furthermore, we assume that, for each $x\in\setX$, there exists $y\in\setY$ with $f(x,y)<+\infty$, implying that $s  \coloneqq \displaystyle \inf_{\substack{ x \in \setX, y \in \setY }} f(x,y) < +\infty$.
Depending on the case, the existence or uniqueness of the minimizer $(x^*,y^*)$ such that $f(x^*,y^*) = s$ may also be assumed. 

The goal of the alternating minimization method is to construct a sequence $\{(x^{(n)},y^{(n)})\}$ such that $\displaystyle \lim_{n \to \infty} f(x^{(n)},y^{(n)}) = s$. Under specific conditions, such a sequence can be defined using the solutions of two sub-problems; for $x_i \in \setX$, $h(x_i) = \displaystyle \argmin_{y \in \setY}  f(x_i,y)$ and, for $y_i \in \setY$, $g(y_i) = \displaystyle \argmin_{x \in \setX}  f(x,y_i)$. Starting from an initial point $y^{(0)}$, we can define the $n$-{th} element of the sequence as:
\begin{align*}
    x^{(n)} = g(y^{(n-1)}) \qquad y^{(n)} = h(x^{(n)}) \qquad \text{for } n = 1,2\ldots.
\end{align*}

Depending on the problem, various sufficient conditions for the existence and optimality of the sequence limit have been studied. For instance,
\begin{itemize}
    \item In \cite{csiszar:1984}, Csiszár and Tusnády prove that, if the sequence $\{(x^{(n)},y^{(n)})\}$ guarantees $\forall x \in \setX, \forall y \in \setY$
    \begin{align*}
        &f(x,y) + f(x,y^{(n-1)}) \ge f(x, y^{(n)}) + f(x^{(n)}, y^{(n - 1)}), 
    \end{align*}
    referred to as \textit{Five-Point property}, then the optimality of the limit is ensured.
    \item In \cite{grippo:2000}, Grippo and Sciandrone prove the convergence of the sequence to a stationary point of $f$, under the assumption of convexity of the feasible sets $\setX$ and $\setY$ and existence of the sequence limit.
\end{itemize}

BA-type algorithms, introduced for the numerical computation of channel capacity \cite{arimoto:72} and RD function \cite{blahut:72}, are specific instances of alternating minimization algorithms \cite[Chapter 9]{yeung:2008}. In fact, in their classic formulation, both problems can be expressed as constrained minimization of a convex function on the sets of marginal distributions and transition matrices, where the properties of the sets (e.g., convexity) depend on the constraints for which the problem is formulated.

\section{Main Results} \label{sec:MainResults}

In this section, we present the derivation of our main theoretical results. We start by providing the parametric characterization of the solution of the RDPF problem in the following lemma, obtained by casting \eqref{opt: primal} as a double minimization problem. 

\begin{lemma} (Double minimization)\label{lemma:DoubleMinimization}
Let $D \ge 0$, $P \ge 0$ and let $D(\cdot||\cdot)=D_f(\cdot||\cdot)$. Moreover, let $s=(s_D, s_P)$ with $s_D\geq{0}$, $s_P\geq{0}$ being the Lagrangian multipliers associated with constraints \eqref{opt:RDPF:DistortionConstraint} and \eqref{opt:RDPF:PerceptionConstraint}. Then \eqref{opt: primal} can be expressed as a double minimum 
\begin{align}
\begin{split}
R(D,P) =  \min_{\substack{\Q{} \in \mathcal{Q}(\setSym) \\ \q{} \in \mathcal{P}(\setSym)} } &D_{KL}(\p{} \Q{} || \p{} \q{} ) \\&+ s_D\left(\E[]{d(X,Y)} - D \right)\\
&+ s_P( D_f(\p{}||\Qm{}) - P)
\end{split}\label{eq: double_min}
\end{align}
where  $D =\E[\Q{*}]{d(X,Y)}$ and $P=D_f(\p{}||\q{*})$, with $(\Q{*},\q{*})$ being the pair achieving the minimum.\\ 
Furthermore, for fixed $\Q{}$, the right-hand side of \eqref{eq: double_min} is minimized by
\begin{align}
\q{}[\Q{}](y) = \sum_{x \in \setSym}  \p{}(x) \Q{}(y|x) \label{eq: optimization qx}
\end{align}
whereas, for fixed $\q{}$, the right-hand side of \eqref{eq: double_min} is minimized by
\begin{align}
\Q{}[\q{}](y|x) &= \frac{\q{}(y) \cdot A[\Qm{}[\q{}]](x,y,s)}{\sum_{i \in \setSym} \q{}(i) \cdot A[\Qm{}[\q{}]](x,i,s)} \label{eq: ParametricQ}
\end{align}
where
\begin{align}
A[u](x,y,s) &= \exp\left\{-s_D d(x,y) - s_P g(\p{}(y), u(y))  \right\} \label{eq: optimization Qxx - A} \\ 
\Qm{}[u](y) &= \ssum{x} \Q{}[u](y|x) \p{}(x) \label{eq: MarginalQ}\\
g(x,y) &=  f\left(\frac{x}{y}\right) - \frac{x}{y} \df\left(\frac{x}{y}\right).\nonumber 
\end{align} 
\end{lemma}
\begin{IEEEproof}
    See Appendix \ref{proof: Minimization Conditions}.
\end{IEEEproof}

We remark that, although showing a close resemblance to the classical BA solution \cite[Theorem 6.3.3]{Blahut:1987}, Lemma \ref{lemma:DoubleMinimization} differs from it in \eqref{eq: ParametricQ}. In particular, the perception constraint \eqref{opt:RDPF:PerceptionConstraint} induces the presence of an additional exponential term, i.e., $s_P  g(\cdot,\cdot)$. Note that the classical BA implicit solution can be obtained as a special case of \eqref{eq: ParametricQ} by considering $s_P = 0$, effectively removing the perceptual constraint.
The next corollary follows as a direct consequence of Lemma \ref{lemma:DoubleMinimization}.
\begin{corollary} \label{cor: Minimization Conditions}
Let $s=(s_D, s_P)$ with $s_D\geq{0}$, $s_P\geq{0}$. Then $R(D,P)$ in \eqref{eq: double_min} can be reformulated as follows
\begin{align}
R(D_s,P_s) = &-s_DD_s -s_PP_s \label{eq: dual after Q optimization} \\ 
+ \min_{\q{} \in \mathcal{P}(\setSym)}& \quad s_P \ssumh{y} \p{}(y) \df\left(\frac{ \p{}(y) }{\Qm{}[\q{}](y)}\right) \nonumber \\ 
&- \ssum{x} \p{}(x)\log\left(\ssumh{y} \q{}(y) A[\Qm{}[h]](x,y,s)\right) \nonumber
\end{align}
where $P_s =  D_f(\p{}||\q{*})$ and
\begin{align}
    D_s &= \dsum{x}{y}  \frac{ \p{}(x) \q{*}(y) A[\q{*}](x,y,s)} {\ssumh{i} \q{*}(i) A[\q{*}](x,i,s) } d(x,y)\nonumber
\end{align}
with $\q{*} \in \mathcal{P}(\setSym)$ achieving the minimum of \eqref{eq: dual after Q optimization}.
\end{corollary}
\begin{IEEEproof}
The proof follows by substitution of \eqref{eq: ParametricQ} into \eqref{eq: double_min}.
\end{IEEEproof}
We note that in Corollary \ref{cor: Minimization Conditions} and in subsequent analysis, the subscript notation $(D_s,P_s)$ is introduced to explicitly indicate the dependence of the constraint levels $(D, P)$ on the fixed Lagrangian multipliers $s = (s_D, s_P)$.
The following lemma characterizes a necessary and sufficient condition to ensure that, for given Lagrangian multipliers $s = (s_D, s_P)$, a distribution $\q{*} \in \mathcal{P}(\setSym)$ is the optimal solution of \eqref{eq: dual after Q optimization}, i.e., $(\q{*}, \Q{}[\q{*}])$ defines a point achieving the RDPF.

\begin{lemma} \label{lemma: opt_condition_q}
Let $D_f(\cdot || \cdot)$ be such that $f \in C^1(0, \infty)$ continuous and differentiable on $(0, \infty)$ and let the vector function $c[\cdot, \cdot]: \mathbb{P}(\setSym)^2 \to \mathbb{R}^{|\setSym|}$ be such that
\begin{align}
    c[u, r](y) = \frac{}{} \ssum{x} \frac{ \p{}(x) A[r](x,y,s)}{\ssumh{i} u(i) A[r](x,i,s)}. \label{eq: c_coeff}
\end{align} 
Then, a probability vector $\q{}$ yields a point on the $R(D,P)$ curve via the transition matrix
$\Q{}$ defined in \eqref{eq: ParametricQ} if and only if $ c[\q{}, \Qm{}[\q{}]](y) \le 1$ for all $y \in \setSym$, holding with equality for any $y$ for which $\q{}(y)$ is nonzero.
\end{lemma}
\begin{IEEEproof}
 See Appendix \ref{proof: opt_condition_q}.
\end{IEEEproof}

\begin{remark}
    It can be shown that the function $c[\cdot,\cdot]$ characterizes also the relation between a distribution $\q{}$ and the result of the functional $\Qm{}[\q{}]$. In fact, we can verify that for all $i \in \setSym$,
    \begin{align*}
        \frac{\Qm{}[\q{}](i)}{\q{}(i)} = \ssum{x} \frac{\Q{}[\q{}](i|x)}{\q{}(i)} \p{}(x) = c[\q{}, \Qm{}[\q{}]](i) .
    \end{align*}
\end{remark}

Using the results of Lemma \ref{lemma:DoubleMinimization}, we now proceed to construct an alternating minimization procedure, thereon referred to as Optimal Alternating Minimization (OAM), proving its convergence to a point of $R(D,P)$. 

\begin{theorem} (OAM) \label{th:OAM}
Let the Lagrangian multipliers $s=(s_D, s_P)$ with $s_D \ge 0$, $s_P \ge 0$  be given. Let $\q{(0)}$ denote any probability vector with nonzero components and let $\Q{(n+1)} \equiv\Q{}[\q{(n)}]$ and $\q{(n+1)} \equiv\Qm{}[\q{(n)}]$ be functions of the current iteration $\q{(n)}$ as defined in \eqref{eq: ParametricQ} and \eqref{eq: MarginalQ}, respectively.
Then, as $n\xrightarrow{}\infty$, we obtain
\begin{align*}
D(\Q{(n)})\xrightarrow{} D_s,~P(\Q{(n)})\xrightarrow{} P_s,~I(\p{}, \Q{(n)}) \xrightarrow{} R(D_s,P_s).
\end{align*}
\end{theorem}
\begin{IEEEproof}
    See Appendix \ref{proof: BA Convergence}.
\end{IEEEproof}

\par Despite being optimal, the OAM scheme does not allow the implementation of a BA-type algorithmic embodiment. The reason stands in the parametric dependencies underlying \eqref{eq: ParametricQ} and \eqref{eq: MarginalQ}, as highlighted in the following remark.

\begin{remark} (Implicit Iterate)
Due to the structure of the iterations in \eqref{eq: ParametricQ} and \eqref{eq: MarginalQ}, an implicit dependency of $\q{(n+1)}$ on itself appears, i.e.,
    \begin{align}
        \frac{\q{(n+1)}(y)}{\q{(n)}(y)} &=\sum_{x \in \setSym} \frac{ \p{}(x) e^{-s_D d(x,y) - s_P g\left( \p{}, {\q{(n+1)}}, y \right) }}{\ssumh{i} \q{(n)}(i) e^{-s_D d(x,i) - s_P g\left( \p{}, {\q{(n+1)}}, i\right)} } \nonumber\\ 
        &= c[\q{(n)},\q{(n+1)}](y)  \label{eq:rec_unfold}
    \end{align}
    showing that the updated term $\q{(n+1)}$ exists in both the left- and the right-hand side of the equation. Consequently, the structure of \eqref{eq:rec_unfold} impedes the characterization of a closed form expression of the "updated term" $\q{(n+1)}$ as a function of only the current iteration term $\q{(n)}$.
\end{remark}

The implementation problem of the OAM scheme prompts us to find alternative ways to compute the alternating minimization iterates. We detail in the following section two different approaches to solve the OAM issue, leveraging either the numerical solution of the implicit equation or through the relaxation of the structure of the iterations. 

\subsection{NAM scheme} \label{sec: NAM}

The implicit definition of $\q{(n+1)}$ in \eqref{eq:rec_unfold} suggests the application of numerical methods for its approximation. To this end, we introduce a variation of the OAM scheme, referred to as NAM scheme, where $\q{(n+1)}$ is approximated at each minimization step using Newton's root-finding method \cite{burden:2015}. 
\par We first demonstrate that the iteration step for $\q{(n+1)}$, i.e., \eqref{eq:rec_unfold}, can be cast as a root finding problem.

\begin{lemma} \label{lemma: Newton-Function Definition}
Let $\q{(n+1)}$ be defined as in Theorem \ref{th:OAM} and let $T:\mathbb{R}^{|\setSym|} \to \mathbb{R}^{|\setSym|}$ be the vector function defined as
\begin{align}
    T[\q{(n)}, u](i) \triangleq u(i) - \q{(n)}(i) \cdot c[\q{(n)},u](i),~ \forall i \in \setSym \label{eq: NewtonRootFunctional}
\end{align} 
where $c[\cdot, \cdot]$ is defined in \eqref{eq: c_coeff}. Then,  $\q{(n+1)}$ is a root of $T[\q{(n)}, \cdot]$, i.e., $T[\q{(n)},\q{(n+1)}] = 0$.
\end{lemma}
\begin{IEEEproof}
    The proof follows from the evaluation of \eqref{eq: NewtonRootFunctional} in $\q{(n+1)}$ and the substitution of \eqref{eq:rec_unfold} therein.
\end{IEEEproof}

The application of Newton's method requires the existence and the invertibility of the Jacobian matrix $J_T$ of the functional $T$ \cite[Section 10.2]{burden:2015}. In our case, ensuring the existence of $J_T$ requires a more restrictive continuity assumption on the divergence measure, i.e., $D_f(\cdot||\cdot)$ needs to be twice differentiable in its second argument. Although this limitation reduces the generality of the NAM scheme, we note that most commonly used divergences (see Section \ref{sec: f-divergences}) satisfy this assumption.
Under this restriction, the invertibility of $J_T$ is shown in the following lemma.

\begin{lemma} \label{lemma: Newton-Jacobian}
    Let $T[\q{(n)}, \cdot]$ be the function defined in \eqref{eq: NewtonRootFunctional} and let the divergence measure $D_f(\cdot||\cdot)$ be twice differentiable in its second argument. Then, the Jacobian $J_T: \mathbb{R}^{|\setSym|} \to \mathbb{R}^{|\setSym| \times |\setSym|} $ of the functional $T[\q{(n)}, \cdot]$, defined as $J_T[\q{(n)}, u] \triangleq \left[ \frac{\partial T[\q{(n)}, v](i)}{\partial v(j)} \Big{|}_{u} \right]_{(i,j) \in \setSym^2}$, is positive definite and has the form
    \begin{align}
        J_T[\q{(n)}, u] = I + \left(C[\q{(n)}, u] - M[\q{(n)}, u]\right) \cdot \Gamma[\q{(n)}, u] \label{eq: NewtonFunctionJacobian}
    \end{align}
    where
    \begin{align}
        &M[\q{(n)}, u]  \label{matrix: Newton_M} = \\
        &~\left[ \q{(n)}(i) \ssum{x} \p{}(x) \frac{\bA[i]{} \cdot \bA[j]{}}{ \left(\ssum{k} \q{(n)}(k) \bA[k]{}\right)^2}  \right]_{(i,j) \in \setSym^2} \nonumber
    \end{align}
    \begin{align} 
        \Gamma[\q{(n)}, u] &=  s_P \diag\left[ \q{(n)}(i) \cdot \frac{\partial^2 D_f(\p{}||v)}{\partial v(i)^2}\bigg{|}_{u} \right]_{i \in \setSym} \label{matrix: Newton_Gamma}\\ 
        C[\q{(n)}, u] &= \diag\Big[c[\q{(n)}, u](i) \Big]_{i \in \setSym}. \label{matrix: Newton_C}
    \end{align}
\end{lemma}
\begin{IEEEproof}
See Appendix \ref{proof: Newton-Jacobian}.
\end{IEEEproof}

We are now ready to define the structure of the iteration of Newton's root-finding method applied to the functional $T(\cdot)$, which, as shown in Lemma \ref{lemma: Newton-Function Definition}, provides an approximation of $\q{(n+1)}$.

\begin{theorem}{(Newton's method)} \label{th: Newton Approximation Method}
  Assume the divergence measure $D_f(\cdot||\cdot)$ to be twice differentiable in its second argument and let $\q{(n+1)}$ and $\q{(n)}$ be defined as in Theorem \ref{th:OAM}. Let $T[\q{(n)}, \cdot]$ and $J_T[\q{(n)}, \cdot]$ be as defined in \eqref{eq: NewtonRootFunctional} and \eqref{eq: NewtonFunctionJacobian}, respectively. Furthermore, let the sequence $\{u^{(k)}\}_{k=1,2,\ldots}$  for some initial point $u^{(0)} \in \mathbb{R}^{|\setSym|} $ be defined as
\begin{align*}
    u^{(k + 1)} \triangleq u^{(k)} - \left(J_T[\q{(n)}, u^{(k)}] \right)^{-1} \cdot T[\q{(n)}, u^{(k)}].
\end{align*}
Then, $\lim_{k \to \infty} u^{(k)} = \q{(n+1)}$.
\end{theorem}
\begin{IEEEproof}
 The proof follows by direct application of Newton's root-finding method \cite[Section 10.2]{burden:2015}, since Lemma \ref{lemma: Newton-Function Definition} proves that the set of solutions $\q{(n+1)}$ and the set of the roots of $T(\cdot)$ coincide, while Lemma \ref{lemma: Newton-Jacobian} proves that $T(\cdot)$ satisfies the assumption for the convergence of the method. 
\end{IEEEproof}

The implementation of the NAM algorithm illustrated in Algorithm \ref{alg: Newton Approximation} is obtained by introducing the results of Theorem \ref{th: Newton Approximation Method} in the OAM scheme defined in Theorem \ref{th:OAM}. However, despite solving the main technical issues of the OAM scheme, the NAM scheme imposes limitations on the choice of the perception metric. In the next section, we provide an alternative minimization scheme that circumvents these issues.

\subsection{RAM scheme} \label{sec: RAM} 
An alternative approach to solve the implementation problems of the OAM scheme is based on a relaxed formulation of the OAM iterations. Through the introduction of an auxiliary design variable $\va{}$ in $\eqref{eq: ParametricQ}$, we define an approximation to the original OAM scheme, referred to as the RAM scheme. The main advantage of the RAM scheme lies in the fact that, for $\va{}$ properly selected, the iterative scheme is directly implementable and does not require additional assumptions on the continuity of the perception constraints, while still being able to achieve a globally optimal solution. The following theorem provides the formal formulation of the RAM iterative scheme.

\begin{theorem}(RAM) \label{th:RAM}  Let the Lagrangian multipliers $s=(s_D, s_P)$ with $s_D \ge 0$, $s_P \in [0, s_{P, \max}]$ be given and define 
\begin{align}
    \tQ{}[u](y|x)  & \triangleq \frac{ u(y) A[\va{}[u]](x,y,s)}{ \ssumh{i} u(i) A[\va{}[u]](x,i,s)} \label{eq:tQ}\\
    \tQm{}[u](y)  & \triangleq \ssum{x} \tQ{}[u](y|x) \p{}(x) \label{eq:tQm}
\end{align}
where $A[\cdot]$ is defined in \eqref{eq: optimization Qxx - A} and $v[\cdot]:\mathcal{P}(\setSym) \to \mathcal{P}(\setSym)$ is any functional defining a probability distribution.
Let $\tq{(0)}$ be any probability vector with nonzero components and let $\tQ{(n+1)} \equiv \tQ{}[\tq{(n)}]$, $\tq{(n+1)} \equiv \tQm{}[\tq{(n)}]$, and $\va{(n)} = v[\tq{(n)}]$. Then, as $n\xrightarrow{}\infty$, we obtain
\begin{align*}
D(\tQ{(n)}) \xrightarrow{} D_s,~P(\tQ{(n)}) &\xrightarrow{} P_s,~I(\p{}, \tQ{(n)}) \xrightarrow{} R(D_s,P_s)
\end{align*}
if $\lim_{n\rightarrow\infty} || \q{(n+1)} - \va{(n)}|| = 0$ with at least linear rate of convergence.
\end{theorem}
\begin{IEEEproof}
See Appendix \ref{proof:BA Convergence Approximate}.
\end{IEEEproof}
Theorem \ref{th:RAM} enables the implementation of the alternating minimization scheme by introducing an auxiliary variable $\va{}[\q{(n)}]$, which approximates the correct iteration $\q{(n+1)}$ while still being a function of only the current iteration of $\q{(n)}$. Nevertheless, depending on $\va{}$, this approximation may incur restrictions on the domain of the Lagrangian multiplier $s_P$ that affect convergence guarantees, as discussed later in Section \ref{sec: Algorithms Analysis}.  

We conclude this section with the following technical remark, highlighting the differences between the NAM and RAM schemes.
\begin{remark} (NAM vs. RAM) 
    The main advantage of NAM is the convergence for any value of the Lagrangian multipliers $(s_D, s_P)$, without the need for any additional condition. However, the introduction of Newton's method requires the differentiability of the perceptual metric and the introduction of additional complexity at each iteration. RAM, on the other hand, removes the differentiability requirement and avoids the additional computational cost at each iteration, but at the expense of a potentially smaller set of $(s_D,s_P)$ for which the algorithm achieves the optimal solution, which may preclude the computation of the complete RDP curve. 
\end{remark}

\section{Algorithmic Implementation and Convergence Analysis} \label{sec: Algorithms Analysis}
This section addresses the algorithmic implementation of the alternating minimization schemes derived in Section \ref{sec:MainResults} and the characterization of their convergence rate.
\par We start by presenting the implementation of the NAM and RAM schemes, respectively, in Algorithm \ref{alg: Newton Approximation} and \ref{alg: RAM}. Subsequently, we discuss the derivation of stopping conditions suitable for both algorithms.

\begin{algorithm}[ht]
    \caption{Newton-based Alternating Minimization (NAM)} \label{alg: Newton Approximation}
    \begin{algorithmic}[1]
        
        \Require source distribution $\p{}$; Lagrangian parameters $s = (s_D, s_P)$ with $s_D \ge 0$ and $s_P \ge 0$; error tolerances $\epsilon>0$, $\delta>0$; distortion measure $d(\cdot,\cdot)$; initial assignment $\q{(0)}$.
        \item[]
        \State  $\omega \gets +\infty$; $n \gets 0$;
        
        \While{$\omega \ge \epsilon$}
            \State $\q{(n+1)} \gets$ \Call{Newton Approx.}{ $\p{}$, $\q{(n)}$, $s$, $\delta$}
            \State $c^{(n)} \gets c \left[\q{(n)},\q{(n+1)} \right]$
            \State $\omega \gets  \log c^{(n)}_{\max}(y) - \ssum{y} \q{(n)} c^{(n)}(y) \log(c^{(n)}(y))$        
            \State $n \gets n + 1$
        \EndWhile

        \item[]
        \Ensure 
        { $D_s = \E[\p{} \tQ{(n)}]{d(X,Y)}$},
        { $ P_s = D_f(\p{}||\q{(n)}) $},
        {  $R(D_{s},P_s) =  \tW[\q{(n)}]-s_DD_s -s_PP_s  - \ssum{y} \q{(n)} c^{(n)} \log(c^{(n)})$, $\tW{}[\cdot]=\eqref{eq: W_tilde_main_text}$. }

    \end{algorithmic}
\end{algorithm}

\begin{algorithm}[ht]
    \caption{Relaxed Alternating Minimization (RAM)} \label{alg: RAM}
    \begin{algorithmic}[1]
        
        \Require source distribution $\p{}$; Lagrangian multipliers $s = (s_D, s_P)$ with $s_D \ge 0$ and $s_P \in [0, s_{P,\max}]$; error tolerance $\epsilon>0$; divergence measure $D_f(\cdot||\cdot)$; distortion measure $d(\cdot,\cdot)$; initial assignment $\tq{(0)}$.
        \item[]
       \State  $\omega \gets +\infty$; $n \gets 0$;
        
        \While{$\omega \ge \epsilon$}
            \State $c^{(n)} \gets c[\tq{(n)},\va{(n)}]$
            \State $\tq{(n+1)} \gets \tq{(n)} \cdot c^{(n)}$
            \State $\omega \gets  \log c^{(n)}_{\max}(y) - \ssumh{y} \tq{(n)} c^{(n)}(y) \log(c^{(n)}(y))$    
            \State $n \gets n + 1$
        \EndWhile
        \item[]
        \Ensure 
        { $D_s = \E[\p{} \tQ{(n)}]{d(X,Y)}$},
        { $ P_s = D_f(\p{}||\tq{(n)}) $},
        { $R(D_{s},P_s) =  \tW[\q{(n)}]-s_DD_s -s_PP_s  - \ssumh{y} \q{(n)} c^{(n)} \log(c^{(n)})$, $\tW{}[\cdot]=\eqref{eq: W_tilde_main_text}$.}

    \end{algorithmic}
\end{algorithm}

\paragraph*{\bf Stopping Criterion} We first derive a stopping criterion for the RAM case, since the NAM case can be obtained by fixing the auxiliary variable $\va{}[\q{(n)}] = \Qm{}[\q{(n)}]$ in Theorem \ref{th:RAM}, i.e., recovering the original OAM iterates. For this purpose, we need the following theorem which establishes bounds on the RDPF. 

\begin{theorem} (Bounds on RDPF)\label{th: UpperAndLowerboundToRate}
Let $\tQ{}$ and $\tQm{}$ be defined  as in Theorem \ref{th:RAM}, $\cx[]$ be as defined as in Lemma \ref{lemma: opt_condition_q}, and $c_{\max} = \max_{y \in \setSym} c[\tq{},\tQm{}[\tq{}]](y)$. Then, at the point $D=\E[\p{}\tQ{}]{d(X,Y)}$, and $P=D_f(\p{}||\tQm{}[\tq{}])$, the following bounds hold
\begin{align} 
    \begin{split}
         R(D,P) & \ge R_L[\tq{}](D,P) \\
        &= -s_DD -s_PP + \tW[\tq{}] - \log(c_{\max})
    \end{split} \label{eq: Lowerbound Approximation}\\
    \begin{split}
        R(D,P)  & \le  R_U[\tq{}](D,P) \\
        &=-s_DD -s_PP + \tW[\tq{}] \label{eq: Upperbound Approximation}\\
        & ~~~~ - \ssumh{y} \tq{}(y) c[\tq{},\tQm{}[\tq{}]](y) \log(c[\tq{},\tQm{}[\tq{}]](y)) 
    \end{split} 
\end{align}
where $\tW{}[\cdot]$ is given by 
\begin{align}
   \begin{split}
    \tW[u] &= - \ssum{x} \p{}(x) \log \left( \ssumh{y} u(y)A[v[u]](x,y,s) \right) \\
    & \quad +s_P \ssumh{y} \tQm{}[u](y) \frac{ \p{}(y) }{v[u](y)} \df \left(\frac{ \p{}(y) }{v[u](y)} \right)\\
    & \quad + s_P \Bigg{[} \ssumh{y} \tQm{}[u] \left(f \left(\frac{ \p{}(y) }{\tQm{}[u](y)} \right) - f \left(\frac{ \p{}(y) }{v[u](y)} \right) \right) \Bigg{]}.
    \end{split} \label{eq: W_tilde_main_text}
\end{align}
\end{theorem}
\begin{IEEEproof}
See Appendix \ref{proof: UpperAndLowerboundToRate}.
\end{IEEEproof}

Leveraging the bounds in \eqref{eq: Lowerbound Approximation} and \eqref{eq: Upperbound Approximation}, we can estimate the precision of the estimation of $R(D,P)$ at the $n$-th iteration by considering the estimation error $\omega = R_U[\tq{(n)}](D,P) - R_L[\tq{(n)}](D,P)$, as implemented in line 5 of both Algorithm \ref{alg: Newton Approximation} and \ref{alg: RAM}.

\subsection{Asymptotic Convergence Rate Analysis}
In this section, we characterize the asymptotic convergence rate of the proposed minimization schemes. We start with the analysis of the convergence rate of the OAM scheme, which, although not directly implementable, serves as a reference for the characterization of the convergence rate of both the NAM and RAM schemes.

\paragraph*{\bf OAM Convergence Rate}
We note that the iteration structure in Theorem \ref{th:OAM}, i.e., $\q{(n+1)}$ = $\Qm{}[\q{(n)}]$ given the current iteration $n$, can be represented as an implicit vector function $S: \realR^{|\setSym|} \to \realR^{|\setSym|}$ with $S[\q{}](i) = \q{}(i) \cdot c[\q{}, S[\q{}]](i)$, such that $\q{(n+1)} = S[\q{(n)}]$. The results of Lemma \ref{lemma: opt_condition_q} characterize a distribution $\q{*}$ that achieves the RDPF as a fixed point of $S(\q{})$, i.e., $\q{*} = S[\q{*}]$, since $c[\q{*}, S[\q{*}]](i) = 1, ~i=1\ldots,|\setSym|$. Under these observations, we can analyze the convergence rate of the OAM scheme following similar steps as in \cite{9476038}. 
\par The first-order Taylor expansion of $S[\q{}]$ around a fixed point $\q{*}$ is defined as
\begin{align*}
 S[\q{}] = S[\q{*}] + J[\q{*}] \cdot (\q{} - \q{*}) + o(||\q{} - \q{*}||)
\end{align*}
where $J[\q{}]$ is the Jacobian matrix of $S[\q{}]$ with entries $
    J[\q{}](i,j) \triangleq \frac{\partial S[\q{}](i)}{\partial \q{}(j)}, (i,j) \in \setSym^2$.
The next theorem provides the functional form of the Jacobian for the case of Theorem \ref{th:OAM}.
\begin{theorem}(Jacobian form) \label{th: base_alg_Jacobian}
The Jacobian $J(\q{})$ evaluated at the fixed point $\q{*}$ is given as 
    \begin{align} 
        J[\q{*}] = \left(I - M^*\right)\left(I - \Gamma^* J[\q{*}]\right) \label{eq: JacobianEq1}
    \end{align}
where $M^* = M[\q{*},\q{*}]$ and $\Gamma^* = \Gamma[\q{*},\q{*}]$ as defined in \eqref{matrix: Newton_M} and \eqref{matrix: Newton_Gamma}, respectively.
\end{theorem}
\begin{IEEEproof}
    See Appendix \ref{proof: base_alg_Jacobian}.
\end{IEEEproof}
Next, we introduce two lemmas, in which we use the structure of \eqref{eq: JacobianEq1} to identify properties of matrix $M^*$.  

\begin{lemma} \label{lemma: lowerboundEigM}
Let $\{ \lambda_i \}_{i \in [1:|\setSym|]}$ be the set of eigenvalues of $M^* = M[\q{*},\q{*}]$. Given a distortion function $d: \setSym \times \mathcal{ Y } \to \realR_0^+$ that induces a full-rank matrix $D=[e^{ - s_D d(i,j)}]_{(i,j) \in \setSym^2}$, then $\lambda_i > 0$,$\forall i \in [1:|\setSym|]$, i.e., $M^*$ has only positive eigenvalues.  
\end{lemma}
\begin{IEEEproof}
    See Appendix \ref{proof: loweboundEigM}.
\end{IEEEproof}

\begin{remark}(On Lemma \ref{lemma: lowerboundEigM}) We note that a popular example that satisfies the assumptions imposed on Lemma \ref{lemma: lowerboundEigM} is the Hamming distortion denoted hereinafter as $d_H$ \cite{cover-thomas:2006}.
\end{remark}

\begin{lemma} \label{lemma: upperboundEigM}
Let $\{ \lambda_i \}_{i \in [1:|\setSym|]}$ be the set of eigenvalues of $M^* = M[\q{*},\q{*}]$. Then, we have that $\lambda_i \le 1, \forall i \in [1:|\setSym|].$
\end{lemma}
\begin{IEEEproof}
    See Appendix \ref{proof: upperboundEigM}.
\end{IEEEproof}

Using Lemmas \ref{lemma: lowerboundEigM} and \ref{lemma: upperboundEigM}, we can characterize the interval that contains the set of eigenvalues of $J[\q{*}]$ and subsequently the convergence rate of Theorem \ref{th:OAM}. 
\begin{theorem}(Convergence rate of Theorem \ref{th:OAM}) \label{th: EXPConvergenceBase}
    Let $\{\theta_i \}_{i \in [1:|\setSym|]}$ be the eigenvalues of $J[\q{*}]$. Then,
    \begin{align*}
        0  \le \{\theta_i \}_{i \in [1:|\setSym|]} < 1.
    \end{align*}
    Moreover, let $\gamma \in [\theta_{\max}, 1)$. Then, there exist $\delta > 0$ and $K > 0$ such that if $\q{(0)} \in \{\q{}: ||\q{} - \q{*}|| \le \delta \}$, we obtain
    \begin{align}
        ||\q{(n)} -  \q{*}|| < K \cdot || \q{(0)} - \q{*}|| \cdot \gamma^n
    \end{align}
    i.e., the iterations converge exponentially.
\end{theorem}
\begin{IEEEproof}
 See Appendix \ref{proof: EXPConvergenceBase}.
\end{IEEEproof}

Summarizing, under the structural constraints on the distortion function $d$ reported in Lemma \ref{lemma: lowerboundEigM}, the exponential convergence of the OAM scheme is guaranteed by Theorem \ref{th: EXPConvergenceBase}.

\paragraph*{\bf NAM Convergence Rate}
The convergence rate of the NAM scheme follows directly from the OAM scheme analysis, given the close relation between the two schemes. Since the only difference lies in the introduction of Newton's root-finding method for the estimation of the optimal iteration step, the NAM scheme exhibits the same convergence rate in terms of the number of iterations as the OAM scheme, i.e., an exponential convergence $\mathcal{O}(e^{-n})$ under the assumptions of Lemma \ref{lemma: lowerboundEigM}. However, the added complexity from the application of Newton's method at each iteration increases the overall iteration complexity, due to the at least linear convergence rate $\mathcal{O}(\frac{1}{m})$ of the root approximant. Therefore, the total complexity is approximately $\mathcal{O}(\frac{e^{-n}}{m})$, where $n$ and $m$ depend on the error tolerances $\epsilon$ and $\delta$ given as input in Algo. \ref{alg: Newton Approximation}.

\paragraph*{\bf RAM Convergence Rate} Following similar steps that led to Theorem \ref{th: EXPConvergenceBase}, the Jacobian $\hat{J}(\q{*})$ associated with the iteration scheme in Theorem \ref{th:RAM} is characterized as
\begin{align*}
    \hat{J}[\q{*}] = (I - M[\q{*},\q{*}])(I - \Gamma[\q{*},\q{*}])
\end{align*}
where $M$ and $\Gamma$ are given by \eqref{matrix: Newton_M} and \eqref{matrix: Newton_Gamma}, respectively. Unlike Theorem \ref{th:OAM}, where the structure of \eqref{eq: JacobianEq1} bounds its own eigenvalues, in this case, we need to bound the Lagrangian multiplier $s_P$, hence matrix $\Gamma$, to guarantee exponential convergence of the algorithm. This is proved in the following theorem.
\begin{theorem} \label{th: approximation_exponential_interval}
    For a given $s_D\ge{0}$, let $I_{s_P} = [0, s_{P,\max}]$ be the domain of $s_P$, $\{\theta_{a,i}\}_{i \in \setSym}$ the set of eigenvalues of $\hat{J}(\q{*})$ and $\theta_{\max}$ the maximum eigenvalue of $J(\q{*})$ in \eqref{eq: JacobianEq1}. Define the set $I_{s_P}^\epsilon = [0, s_{P,\max} - \epsilon]$ for $0 < \epsilon < s_{P,\max} $. Then, there exists an $\epsilon^\prime$ such that if $s_P \in I_{s_P}^{\epsilon^\prime}$ then $0 \le \{\theta_{a,i}\}_{i \in \setSym} < 1$. 
\end{theorem}
\begin{IEEEproof}
    See Appendix \ref{proof: approximation_exponential_interval}.
\end{IEEEproof}
Theorem \ref{th: approximation_exponential_interval} guarantees exponential convergence for Theorem \ref{th:RAM} only for $s_P \in  I_{s_P}^{\epsilon}$ which means that we can consider $P \in [P_{\min}(\epsilon), P_{\max}]$, depending on the characteristics of the input in a specific problem. 

\section{Numerical Results} \label{sec: Numerical Examples}
In this section, we validate our theoretical findings through simulations. In particular, we consider the computation of the RDPF under the NAM and RAM schemes, using Algorithms \ref{alg: Newton Approximation} and \ref{alg: RAM}, respectively.
\subsection{RDPF Computation - NAM scheme}
\begin{figure}[!t]
    \centering
    \subfloat[$D_{JS}(\cdot||\cdot)$]{%
        \includegraphics[width=\linewidth]{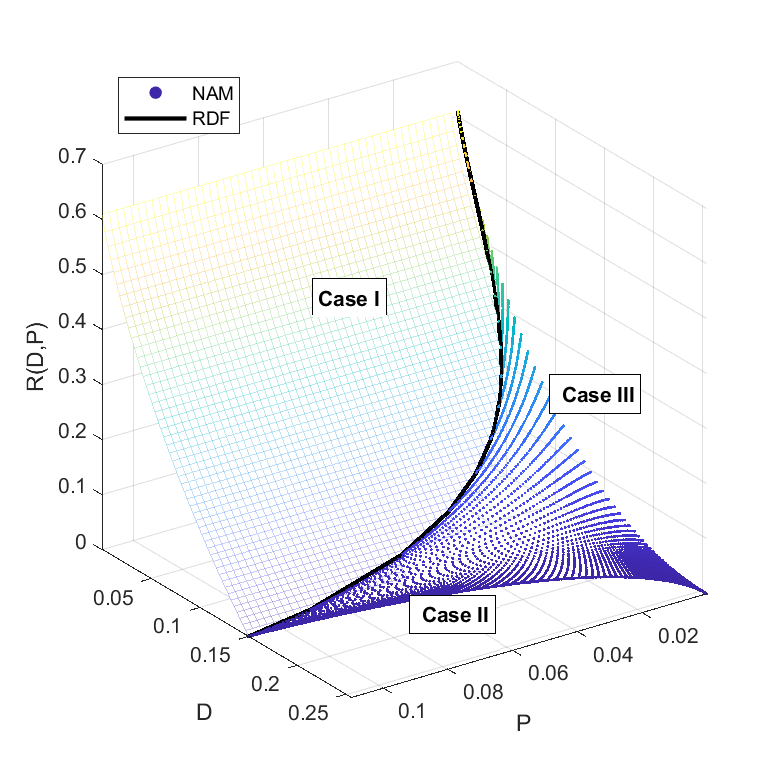}
        \label{fig:NAMCasesExample}
    }

    \subfloat[$D_{KL}(\cdot||\cdot)$]{%
        \includegraphics[width=0.47\linewidth]{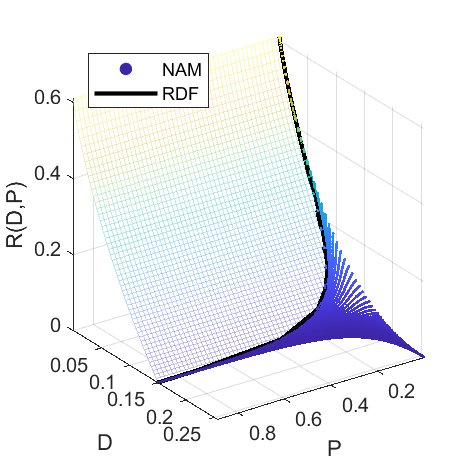}
    }
    \hfill
    \subfloat[$D_{\chi^2}(\cdot||\cdot)$]{%
        \includegraphics[width=0.47\linewidth]{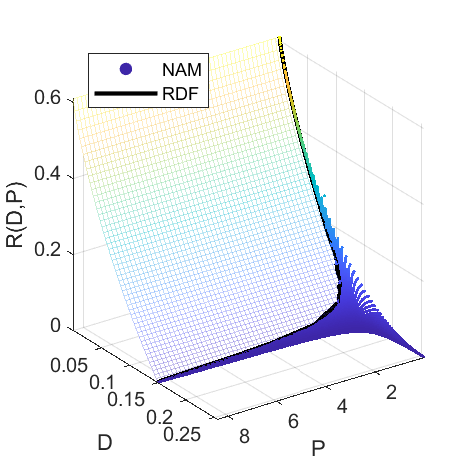}
    }

    \subfloat[$D_{\alpha = -1}$]{%
        \includegraphics[width=0.47\linewidth]{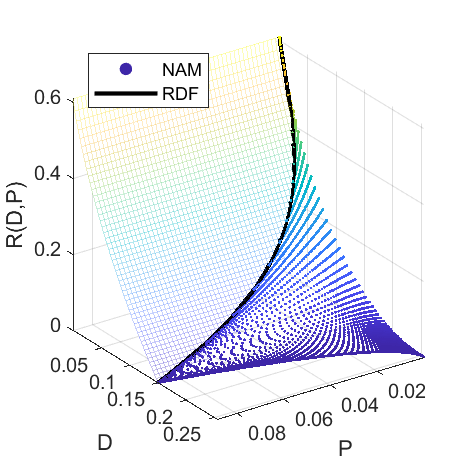}
    }
    \hfill
    \subfloat[$D_{\alpha = \frac{1}{2}}$]{%
        \includegraphics[width=0.47\linewidth]{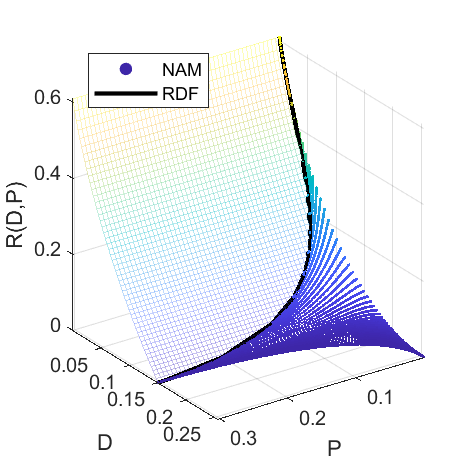}
    }

    \caption{$R(D,P)$ for a Bernoulli source under Hamming distortion and various perception constraints.}
    \label{fig:NAM_RDP}
\end{figure}

\begin{example} \label{example: NAM}
Suppose that ${\mathcal X} = \{0,1\}$ with $\p{} \sim Ber(0.15)$, and let $d(\cdot,\cdot) = d_H(\cdot,\cdot)$ with the perception constraint chosen as one of the following: (a) $ D_f(\cdot||\cdot) = D_{JS}(\cdot||\cdot)$, (b) $D_f(\cdot||\cdot) = D_{KL}(\cdot||\cdot)$, (c) $D_f(\cdot||\cdot) = D_{\chi^2}(\cdot||\cdot)$, (d) $ D_f(\cdot||\cdot) = D_{\alpha=-1}(\cdot||\cdot)$, (e) 
$D_f(\cdot||\cdot)=D_{\alpha=\frac{1}{2}}(\cdot||\cdot)$. In Fig. \ref{fig:NAM_RDP}, we present the $R(D,P)$ estimates obtained using Algorithm \ref{alg: Newton Approximation} for each divergence metric.
\end{example}

We observe that all the computed RDPFs share similarities in the structure of the operating regions on the $(D,P)$ plane. Referring to the plot of Fig. \ref{fig:NAMCasesExample}, we distinguish three cases:
\begin{itemize}
    \item {\it Case I}, where the perception constraint $P$ is not met with equality. In this region, the RDPF is equal to the RDF.
    \item {\it Case II}, where, due to the distortion level $D$ not met with equality, the RDPF function is identically zero.
    \item {\it Case III}, where both the distortion level $D$ and perception level $P$ are met with equality.
\end{itemize}
The results show that Algorithm \ref{alg: Newton Approximation} can completely cover the {\it Case III} region. The boundary between {Cases I-III} (black line) representing the RDF is obtained by setting the Lagrangian multipliers $s_P = 0$. Furthermore, the {\it Case I} region can be obtained by extension of the relative boundaries, since all the $(D,P)$ points share the solution given by the RD problem.

\subsection{RDPF Computation - RAM scheme}
\begin{figure}[!t]
    \centering
    \subfloat[$D_{JS}(\cdot||\cdot)$]{%
        \includegraphics[width=\linewidth]{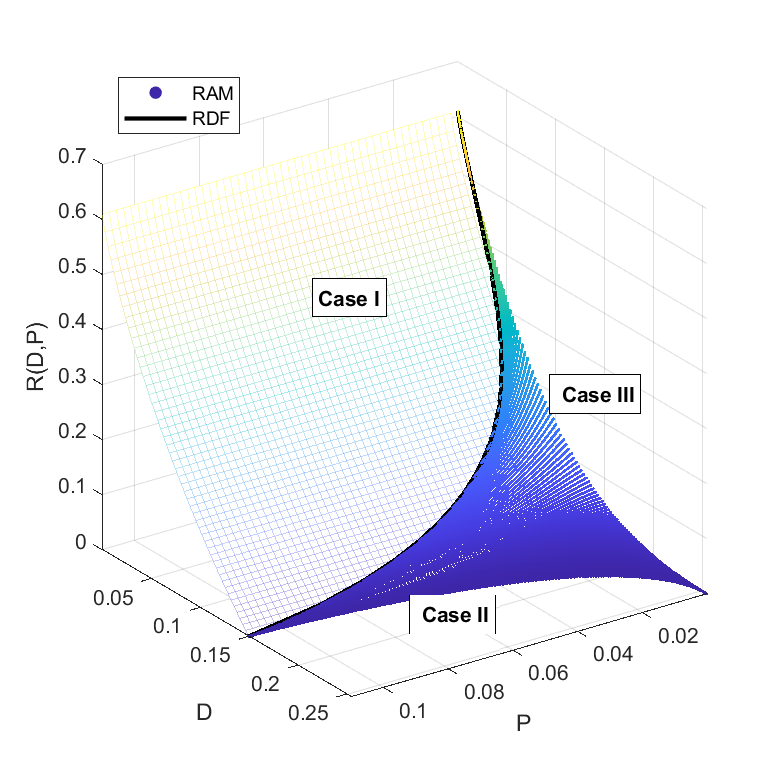}
        \label{fig:RAM_JS}
    }

    \subfloat[$D_{KL}(\cdot||\cdot)$]{%
        \includegraphics[width=0.47\linewidth]{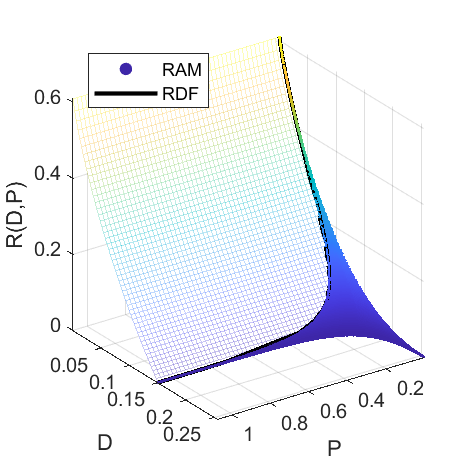}
        \label{fig:RAM_KL}
    }
    \hfill
    \subfloat[$D_{\chi^2}(\cdot||\cdot)$]{%
        \includegraphics[width=0.47\linewidth]{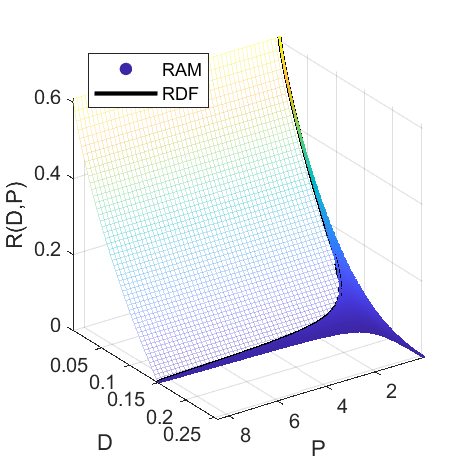}
        \label{fig:RAM_CHI2}
    }

    \subfloat[$D_{\alpha = -1}(\cdot||\cdot)$]{%
        \includegraphics[width=0.47\linewidth]{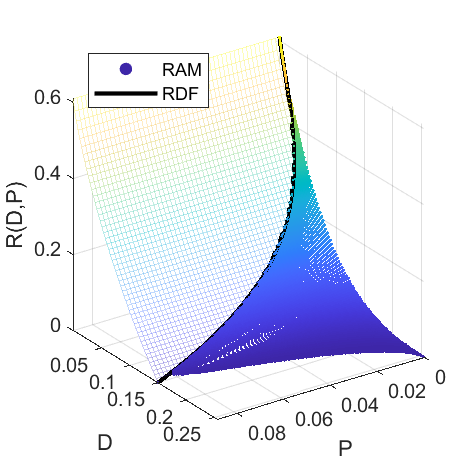}
        \label{fig:RAM_Alpha-1}
    }
    \hfill
    \subfloat[$TV(\cdot||\cdot)$]{%
        \includegraphics[width=0.47\linewidth]{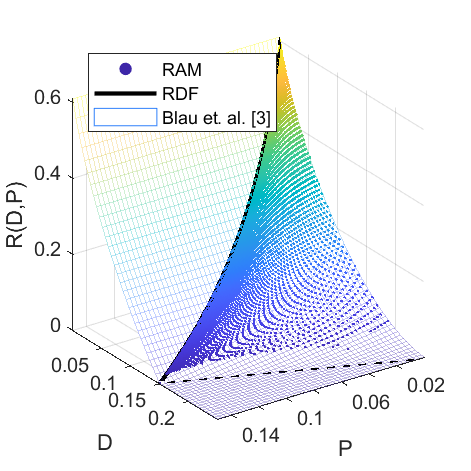}
        \label{fig:RAM_TV}
    }

    \caption{$R(D,P)$ for a Bernoulli source under Hamming distortion and various perception constraints.}
    \label{fig:RAM_RDP}
\end{figure}

\begin{example} \label{example: RAM}
Suppose that ${\mathcal X}=\hat{\cal X}=\{0,1\}$ with $\p{} \sim Ber(0.15)$, and let $d(\cdot,\cdot)=d_H(\cdot,\cdot)$ with perception constraint chosen to be either (a) $ D_f(\p{}||\q{}) = D_{JS}(\cdot||\cdot)$, (b) $D_f(\cdot||\cdot) = D_{KL}(\cdot||\cdot)$, (c) $D_f(\cdot||\cdot) = D_{\chi^2}(\cdot||\cdot)$, (d) $ D_f(\cdot||\cdot) = D_{\alpha=-1}(\cdot||\cdot)$, (e) 
$D_f(\cdot||\cdot) = TV(\cdot||\cdot)$. In Fig. \ref{fig:RAM_RDP}, we present the $R(D,P)$ estimates obtained using Algorithm \ref{alg: RAM} for each divergence metric.
\end{example}

Similar remarks to those in Example \ref{example: NAM} apply to the operating regions of the computed RDPFs. In addition, in Fig. \ref{fig:RAM_TV} we compare the theoretical results of \cite[Equation 6]{blau:2019} with the numerical results obtained using Algorithm \ref{alg: RAM}. This case is of particular interest due to the non-differentiability of the $TV$ distance. We observe that Algorithm \ref{alg: RAM} achieves exactly the theoretical solution of \cite[Equation 6]{blau:2019} as long as $D\le{D}_{\max}=0.15$. We attribute this behavior to limitations in the domain of the Lagrangian multiplier $s_P$, as discussed in Theorem \ref{th: approximation_exponential_interval}. We address this issue in detail in the following section.\\

\subsection{On the convergence under TV perception constraint} \label{sec: OnTheTV}
As observed in Example \ref{example: RAM}, in the case of TV distance, the RAM algorithm is not able to converge to proper solutions in the region $\Omega = \{ (D,P)\in \realR_+^2: D \ge p = 0.15 \}$. We argue that the problem stems from the values of the Lagrangian multipliers $s = (s_D,s_P)$ associated with $\Omega$ require values of $s_P$ that do not guarantee the convergence of the algorithm, as reported in Theorem \ref{th: approximation_exponential_interval}. 
\par To solve the issue, we propose an approximation of the $TV(\cdot||\cdot)$ distance through a sequence of $f$-divergences $\{D_{f_n}\}_{n \in \mathbb{N}}$ such that ${D_{f_n}} \to TV$ for $n \to \infty$. We start with the following general property.

\begin{lemma} \label{lemma: RDP_Perception_upperbound}
    For a divergence metric $D_f(\cdot||\cdot)$, let the set $\mathcal{L}_{D_f}(D,P)$ be defined as
    \begin{align*}
        \mathcal{L}_{D_f}(D,P) \triangleq \{ \Q{}: \E[\Q{}]{d(X,Y)} \le D , D_f(\p{}|| \Qm{}) \le P \}.
    \end{align*}
    Given $D_{f}$, $D_{g}$ divergence metrics with $D_{f}(p||q) \le D_{g}(p||q) , \forall p,q \in \mathcal{P}$, then $\mathcal{L}_{D_{g}}(D,P) \subseteq \mathcal{L}_{D_{f}}(D,P)$. Moreover, for the associated RDPF problems
    \begin{align*}
        R_{D_{f}}(D,P) = \min_{\Q{} \in \mathcal{L}_{D_{f}}(D,P)} I(\p{}, \Q{}) \\ R_{D_{g}}(D,P) = \min_{\Q{} \in \mathcal{L}_{D_{g}}(D,P)} I(\p{}, \Q{})
    \end{align*}
     the inequality $R_{D_{g}}(D,P) \ge R_{D_{f}}(D,P)$ holds.
\end{lemma}
\begin{IEEEproof}
    The inequality $R_{D_{g}}(D,P) \ge R_{D_{f}}(D,P)$ holds if $\mathcal{L}_{D_g} \subseteq \mathcal{L}_{D_{f}}$, which is a trivial implication of $D_{f}(p||q) \le D_{g}(p||q), \forall p,q \in \mathcal{P}$.
\end{IEEEproof}

We can now characterize a sequence $\{D_{f_n}\}$ such that  ${D_{f_n}} \to TV$ for $n \to \infty$ and ${D_{f_n}} \le TV$.

\begin{lemma} \label{lemma: LowboundTV}
    Let ${f_n}$ be the sequence of convex functions defined as $f_n(x) = \frac{2}{\pi}(x-1) \arctan(n(x-1))$ and let $\{D_{f_n}(\cdot||\cdot)\}_{n = 1,2,\ldots}$ be the sequence of associated $f$-divergences. Then, for $n\to \infty$, $D_{f_n} \to TV$ uniformly. Furthermore, for $n  = 1,2, \ldots$ and for all $p,q \in \mathcal{P}(\setSym)$, $ D_{f_n}(p||q) \le TV(p||q)$.
\end{lemma}
\begin{IEEEproof}
    See Appendix \ref{proof: LowboundTV}.
\end{IEEEproof}

\begin{figure}[!t]
    \centering
    \subfloat[$TV(\cdot||\cdot)$]{%
        \includegraphics[width=0.47\linewidth]{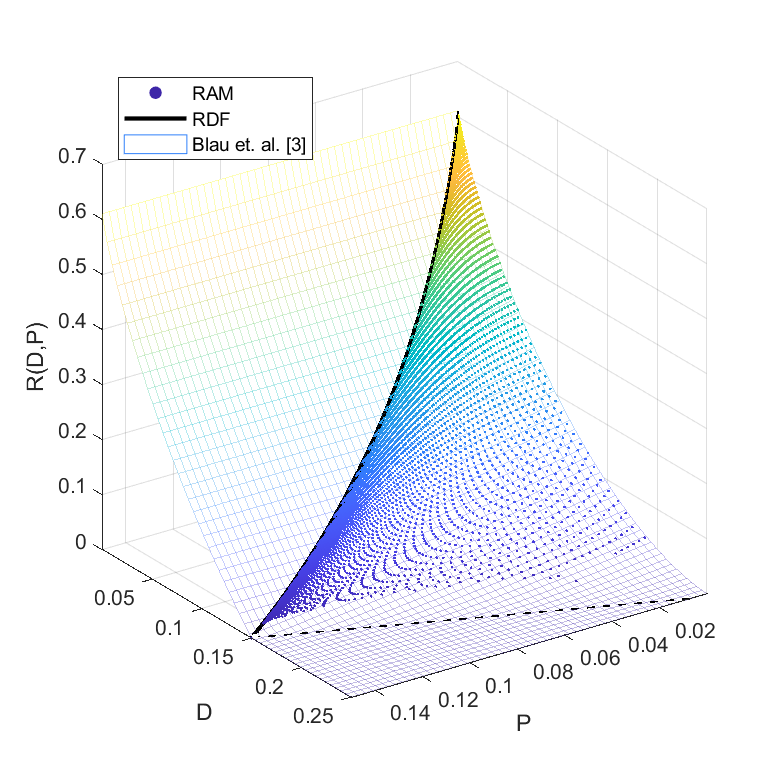}
        \label{fig:TV}
    }
    \hfill
    \subfloat[$n = 1$]{%
        \includegraphics[width=0.47\linewidth]{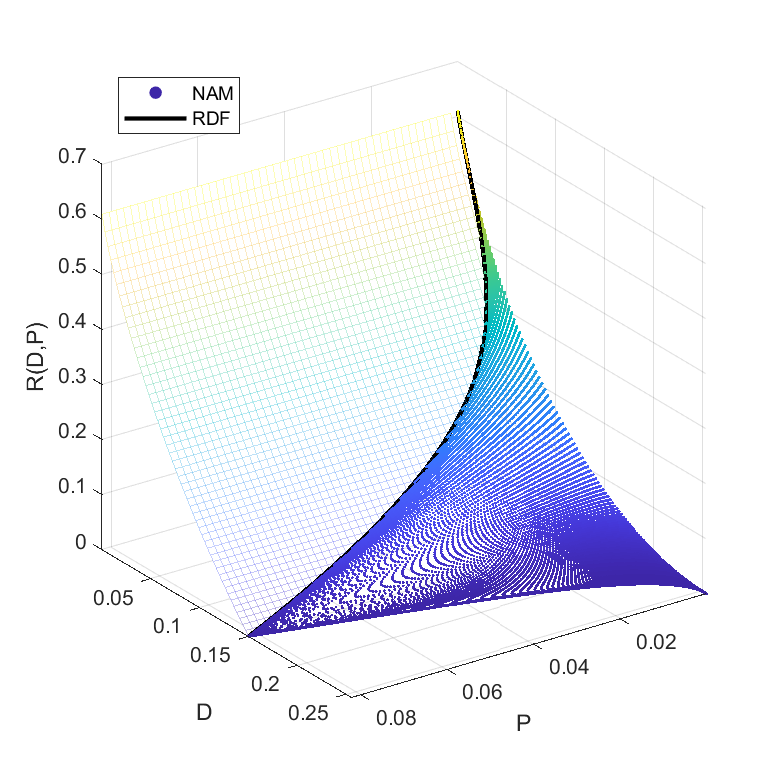}
        \label{fig:Tanh1}
    }

    \subfloat[$n = 10$]{%
        \includegraphics[width=0.47\linewidth]{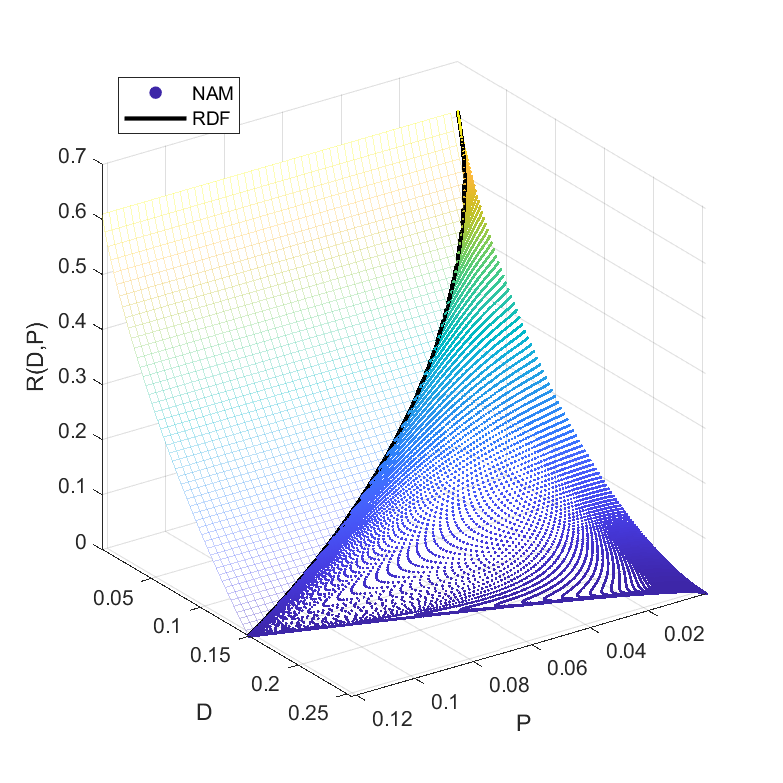}
        \label{fig:Tanh10}
    }
    \hfill
    \subfloat[$n = 100$]{%
        \includegraphics[width=0.47\linewidth]{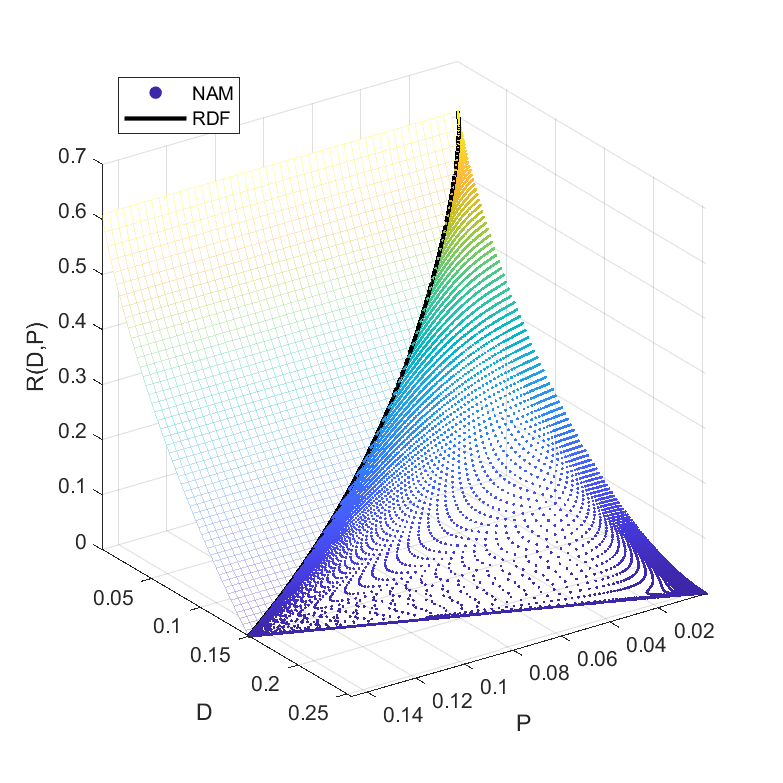}
        \label{fig:Tanh100}
    }

    \caption{Comparison between the RDPF under $TV$ perception, computed with the RAM scheme, and the RDPF under $D_{f_n}$, computed with the NAM scheme, for $n \in \{1, 10, 100\}$.}
    \label{fig:ComparisonTV}
\end{figure}

The results from Lemmas \ref{lemma: RDP_Perception_upperbound} and \ref{lemma: LowboundTV} guarantee that, $\forall n \in \mathbb{N}$, the RDP problem defined for the perception metric $D_{f_n}(\cdot||\cdot)$ acts as a lower bound for the RDP problem defined with $TV(\cdot||\cdot)$ perception metric. 
Furthermore, since $D_{f_n}(\cdot||\cdot)$ is a smooth function, we can apply the NAM scheme to compute the associated RDPF.
Fig. \ref{fig:ComparisonTV} shows the RDPFs for a source $X \sim Ber(0.15)$ under Hamming distortion and perception measure $D_{f_n}$, for $n \in \{ 1, 10, 100 \}$. As expected, for increasing $n$, $D_{f_n}$ provides a progressively better approximation of the $TV(\cdot||\cdot)$ metric.

\section{Conclusion and Future Directions}
In this paper, we have studied the computation of the rate-distortion-perception function for discrete memoryless
sources subject to a single-letter average distortion constraint and a perception constraint from the family of $f$-divergences. We showed that the introduction of the perception constraint induces a non-trivial generalization of the classical optimality conditions for the reconstruction distribution and the transition matrix $(\q{*},\Q{*})$, resulting in the OAM scheme. Since the OAM scheme involves implicit equations that preclude direct algorithmic implementation, we introduced alternative minimization schemes, NAM and RAM, each with its own applicability conditions. Furthermore, we analyzed the asymptotic complexity of the proposed schemes and provided numerical results to validate our theoretical findings. \\
As ongoing research, we plan to extend the derived optimality conditions to general measure spaces, as well as to study the behavior of the proposed algorithms when applied to continuous sources.

\appendices

\section{Proof of Lemma \ref{lemma:DoubleMinimization}} \label{proof: Minimization Conditions}
The double minimization follows immediately from properties of the KL divergence, reported in \cite[Theorem 5.2.6]{Blahut:1987}, which allows to rewrite \eqref{opt: primal} as \eqref{eq: double_min}. 
Moreover, the characterization \eqref{eq: optimization qx} of the minimizer $\q{}$ follows from the same properties.
\par On the other hand, we remark that \eqref{opt: primal} is a convex program in the variable $ \Q{} $ for a given $p_X$, and respects Slater's condition, since it is easy to show that $\Q{}(y|x) = \delta_{y = x}$ is an interior point of the constraint set, as highlighted in Remark \ref{remark:2}. Therefore, the minimizer $ \Q{*} $ can be characterized by applying the Karush-Kuhn-Tucker (KKT) conditions \cite{rockafellar1970convex} on the dual formulation of \eqref{opt: primal}. 

The Lagrangian associated with the primal problem has the form:
\begin{align*}
    & L(\Q{}, s, \lambda, \mu) \\
    &= D_{KL} \left( \p{}\Q{} || \p{} \q{} \right)  + s_D \left( \E{d(X,Y)} - D \right) \\ 
    & \quad + s_P \left(D_f(\p{}||\Qm{}) - P \right) + \ssum{x} \lambda_x \left( 1 - \ssumh{y} \Q{}(y|x) \right) \\
    & \quad -\dsum{x}{y} \mu_{x,y}\Q{}(y|x) 
\end{align*}
where the last two sets of constraints refer to the positivity and normalization constraints on $\Q{}$. Differentiating $L(\Q{}, s, \lambda, \mu)$ with respect to the primal variables $\Q{}(y|x), \forall (x,y) \in \setSym^2$, we obtain
\begin{align*}
    \begin{split}
        &\frac{\partial L(\Q{}, s, \lambda, \mu) }{ \partial\Q{}(y|x)} \\ & ~~~~~~~~~=  \frac{\partial D_{KL}(\p{}\Q{} || \p{} \q{})}{\partial\Q{}(y|x)} + s_D \frac{\partial E[d(X,Y)]}{ \partial\Q{}(y|x)} \\& ~~~~~~~~~~~~~+s_P \frac{\partial D_f(\p{}||\Qm{}) }{ \partial\Q{}(y|x)} + \lambda_x - \mu_{x,y}
    \end{split}
\end{align*}
where:
\begin{align*}
    \begin{split}
    \frac{\partial D_{KL}(\p{}\Q{} || \p{} \q{}) }{ \partial\Q{}(y|x)} &= \p{}(x) \left( \log(\frac{Q_{Y|X}(y|x)}{\q{}(y)}) + 1 \right)   
    \end{split}\\
    \begin{split}
    \frac{\partial E[d(X,Y)]}{ \partial\Q{}(y|x)} &= \p{}(x) d(x,y)
    \end{split}\\
    \begin{split}
    \frac{\partial D_f(\p{}||\Qm{}) }{ \partial\Q{}(y|x)} &=  \p{}(x) g(\p{}, \Qm{}, y).
    \end{split}
\end{align*}
Enforcing \textit{stationarity} and \textit{complementary slackness} 
\begin{align} \label{eq:KKT Conditions}
\begin{cases}
    \begin{split}
        \frac{\partial L(\Q{}, s, \lambda, \mu) }{ \partial\Q{}(y|x)} = 0 \\
        \mu(x,y)\Q{}(y|x)  = 0 \\
    \end{split}
\end{cases}
\end{align}
we solve for $Q_{Y|X}(y|x)$ while choosing $\lambda(x)$ such that $\sum_{y\in \setSym}\Q{}(y|x) = 1$, $\forall x \in \setSym$, resulting in \eqref{eq: ParametricQ} and \eqref{eq: optimization Qxx - A}. This completes the proof.

\section{Proof of Lemma\ref{lemma: opt_condition_q}} \label{proof: opt_condition_q}

Using the results of Corollary \ref{cor: Minimization Conditions}, we can apply KKT conditions to (\ref{eq: dual after Q optimization}). Thus, a minimum for $\q{}$ must satisfy:
\begin{align*}
    \begin{split}
           \frac{\partial}{\q{}(y)}  &\Bigg[ - \sum_{x \in  \setSym} \p{}(x)\log \left(\ssumh{y} \q{}(y) A[h](x,y,s) \right) \\ &+ s_P \ssumh{y} \p{}(y) \df \left(\frac{ \p{}(y) }{\Qm{}[\q{}](y)} \right) + \lambda \ssumh{y} \q{}(y) \Bigg] \ge 0
    \end{split}
\end{align*}
which reduces to
\begin{align*}
    \lambda - c[h,\Qm{}[h]](y)  \ge 0.
\end{align*}

The Lagrangian multiplier $\lambda$ is evaluated by multiplying by $\q{}$ and summing over $y \in \setSym$, giving $\lambda = 1$. This concludes the proof.

\section{Proof of Theorem \ref{th:OAM}} \label{proof: BA Convergence}

Let $\textbf{h}=(\q{(0)}, \q{(1)}, \ldots)$, and $\textbf{Q}=(\Q{(1)},\Q{(2)}, \ldots)$ be the sequences of probability vectors and transition matrices obtained by the chain of alternating minimization $\q{(n)} \xrightarrow{} \Q{(n+1)} \xrightarrow{} \q{(n+1)}$. Let $A^{(n)} = A[\q{(n+1)}]$ and
define the functionals $V[\q{(n)}]$ and $W[\q{(n)}]$ as
\begin{align}
    V[\q{(n)}] = & D_{KL}\left(\p{} \cdot \Q{(n+1)}|| \p{} \cdot \q{(n)}\right) \label{eq: V}\\ 
    &+ s_D\E[\p{}\Q{(n+1)}]{d(X,Y)} + s_P D_f(\p{}|| \q{(n+1)}) \nonumber\\
    W[\q{(n)}] &= s_P \sum_{y \in  \setSym} \p{}(y) \df\left(\frac{ \p{}(y) }{\q{(n+1)}(y)}\right) \label{eq: W}\\
    & ~~~~- \sum_{x \in  \setSym} \p{}(x)\log\left(\sum_{y \in  \setSym} \q{(n)}(y)\A{(n)}\right) \nonumber.
\end{align}
 Using Theorem \ref{lemma:DoubleMinimization} and Corollary \ref{cor: Minimization Conditions}, we can show that $V[\textbf{q}]$ is non-increasing on the sequence $\textbf{q}$ by first fixing $\q{(n)}$ and minimizing over $\textbf{Q}$ (resulting in $\Q{(n+1)}$) and then fixing $\Q{(n+1)}$ and minimizing over $\textbf{q}$ (resulting in $\q{(n+1)}$). $W[\q{(n+1)}]$ is the result of the minimization over $\textbf{Q}$ between $V[\q{(n)}]$ and $V[\q{(n+1)}]$, resulting in a non-increasing sequence
    $V[\q{(n)}] \ge W[\q{(n)}] \ge V[\q{(n+1)}]$.
Given $V[\textbf{q}]$ non-increasing and bounded from below, $V[\textbf{q}]$ converges to some number $V^\infty$.

\par Let now $\q{*}$ be any probability vector and $\Q{*} = \Q{}[\q{*}]$ such that $  R(D,P) = V[\q{*}] - s_DD - s_PP$ and let 
\begin{align}
\begin{split}
&\dsum{x}{y} \p{} \Q{*} \log\left(\frac{\Q{(n)}}{\Q{(n+1)}}\right) \\&=  \quad\dsum{x}{y} \p{} \Q{*} \log\left(\frac{\Q{(n)}}{\q{(n)}}\right) \\ 
&\qquad- \dsum{x}{y} \p{} \Q{*} \log\left(\A{(n)}\right)\\
&\qquad + \dsum{x}{y} \p{} \Q{*} \log\left(\ssumh{i} \q{(n)}(i) \A[i]{\left(n\right)}\right)
\end{split} \label{eq: tmp1}
\end{align}
where
\begin{align*}
&\dsum{x}{y} \p{} \Q{*} \log\left(\A{(n)}\right) \\
& \quad = -s_D\E[\Q{*}]{d(X,Y)} - s_P \ssumh{y} \q{*}(y) f\left(\frac{ \p{}(y) }{\q{(n+1)}(y)}\right) \\ 
& \qquad +s_P \ssumh{y} \q{*}(y) \frac{ \p{}(y) }{\q{(n+1)}(y)} \df\left(\frac{ \p{}(y) }{\q{(n+1)}(y)}\right).
\end{align*}
We can introduce an upper bound to \eqref{eq: tmp1} by noticing that
\begin{align}
    \begin{split}
        &\dsum{x}{y} \p{} \Q{*} \log\left(\frac{\Q{(n)}}{\q{(n)}}\right) -  \p{} \Q{*} \log\left(\frac{\Q{*}}{\q{*}}\right)\\
        & \quad = \dsum{x}{y} \p{} \Q{*} \log\left(\frac{\Q{(n)} \cdot \q{*} }{\q{(n)} \cdot \Q{*} }\right) \\
        & \quad \le \dsum{x}{y} \p{} \Q{(n)} \log\left(\frac{\q{*} }{\q{(n)}}\right) - 1 = 0.
    \end{split} \label{eq: utility_derivation_1}
\end{align}
As a result, we obtain
\begin{align*}
&\dsum{x}{y} \p{} \Q{*} \log\left(\frac{\Q{(n)}}{\Q{(n+1)}}\right) \\
& \le \dsum{x}{y} \p{} \Q{*} \log\left(\frac{\Q{*}}{\q{*}}\right) \\
& \quad + s_D\E[\Q{*}]{d(X,Y)} + s_P D_f(\p{}||\q{*})\\
& \quad +s_P\left[\ssumh{y} \q{*}(y) f\left(\frac{ \p{}(y) }{\q{(n+1)}(y)}\right) - D_f\left(\p{}||\q{*}\right)\right]\\
& \quad - \Biggl[ -\dsum{x}{y} \p{} \Q{*} \log\left(\ssumh{i} \q{(n)}(i) \A[i]{(n)}\right) \\
& \qquad + s_P \ssumh{y} \q{(n+1)}(y) \frac{ \p{}(y) }{\q{(n+1)}(y)} \df\left(\frac{ \p{}(y) }{\q{(n+1)}(y)}\right)\Biggr]\\
& \quad +s_P \Biggl[\ssumh{y} \q{(n+1)}(y) \frac{ \p{}(y) }{\q{(n+1)}(y)} \df\left(\frac{ \p{}(y) }{\q{(n+1)}(y)}\right) \\
& \qquad - \ssumh{y} \q{*}(y) \frac{ \p{}(y) }{\q{(n+1)}(y)} \df\left(\frac{ \p{}(y) }{\q{(n+1)}(y)}\right)\Biggr]\\
& \le V[\q{*}] - W[\q{(n)}] \\
& \quad +s_P\left[\ssumh{y} \q{*}(y) f\left(\frac{ \p{}(y) }{\q{(n+1)}(y)}\right) - D_f(\p{}||\q{*})\right]\\
& \quad +s_P\Biggl[\ssumh{y} \q{(n+1)}(y) \frac{ \p{}(y) }{\q{(n+1)}(y)} \df\left(\frac{ \p{}(y) }{\q{(n+1)}(y)}\right) \\
&\qquad - \ssumh{y} \q{*}(y) \frac{ \p{}(y) }{\q{(n+1)}(y)} \df\left(\frac{ \p{}(y) }{\q{(n+1)}(y)}\right)\Biggr].
\end{align*}
Notice that for any iteration $n\ge0$, we have $ W[\q{(n)}] \ge V[\q{(n+1)}] \ge V[\q{*}] $ and, subsequently, the following inequalities
\begin{align}       
0 \le W[\q{(n)}] - V[\q{*}] \le G[\q{(n)}]\label{tmp2} 
\end{align}
where
\begin{align*}  
G[\q{(n)}] & = \dsum{x}{y} \p{} \Q{*} \log \left(\frac{\Q{(n+1)}}{\Q{(n)}} \right) \\
&  \quad +s_P \left\{ \E[\q{*}]{f \left(\frac{ \p{}}{\q{(n+1)}} \right)} - D_f \left(\p{}||\q{*} \right) \right\}\\
& \quad +s_P \Biggl\{ \E[\q{(n+1)}]{ \frac{ \p{} }{\q{(n+1)}} \df \left(\frac{\p{}}{\q{(n+1)}} \right)} \\
& \qquad ~~~~~~~-\E[\q{*}]{ \frac{ \p{} }{\q{(n+1)}} \df\left(\frac{\p{}}{\q{(n+1)}}\right)} \Biggr\}.
\end{align*}
Since $s_P \ge 0$ and due to the fact that
\begin{align*}
& \E[\q{*}]{f \left(\frac{ \p{}}{\q{(n+1)}} \right)} - D_f \left(\p{}||\q{*} \right) \\
& \quad +\E[\q{(n+1)}]{ \frac{ \p{} }{\q{(n+1)}} \df \left(\frac{\p{}}{\q{(n+1)}} \right)} \\
& \quad -\E[\q{*}]{ \frac{ \p{} }{\q{(n+1)}} \df \left(\frac{\p{}}{\q{(n+1)}} \right)}\\
& = D_f(\p{}||\q{(n+1)}) - D_f(\p{}||\q{(*)}) \\
&\quad - \ssum{i} \left(\q{*}(i) - \q{(n+1)}(i)\right) f \left(\frac{\p{}(i)}{\q{(n+1)}} \right) \\
&\quad + \ssum{i} \left(\q{*}(i) - \q{(n+1)}(i)\right) \frac{\p{}(i)}{\q{(n+1)}} \df \left(\frac{\p{}(i)}{\q{(n+1)}} \right)\\
&= D_f\left(\p{}||\q{(n+1)}\right) - D_f\left(\p{}||\q{(*)}\right) \\
& \quad - \ssum{i} \left( \q{(n+1}(i) - \q{*}(i) \right) \frac{\partial D_f(\p{}||u)}{\partial u(i)}\Bigg{|}_{\q{(n+1)}}\\
&= -\left( D_f \left(\p{}||\q{*}\right) - T_{D_f(p||\cdot),\q{(n+1)}}(\q{*}) \right) \stackrel{(a)}\le 0  
\end{align*}
where $T_{D_f(p||\cdot),\q{(n+1)}}(\q{*}))$ is the first order Taylor expansion of $D_f(\p{}||\cdot)$, centered in $\q{(n+1)}$ and evaluated in $\q{(*)}$,  and $(a)$ is verified since $D_f(\p{}||\cdot)$ is a convex function in its second argument. Therefore $\forall n \in \mathbb{N}$, we obtain
\begin{align}        
    W[\q{(n)}] - V[\q{*}] \le & \dsum{x}{y} \p{} \Q{*} \log \left( \frac{\Q{(n+1)}}{\Q{(n)}} \right). \nonumber
\end{align} 
Summing over $N$ terms, we obtain
\begin{align*}
&\sum_{n = 1}^{N} (W[\q{(n)}]- V[\q{*}])\\
&\quad \le\dsum{x}{y} \p{} \Q{*} \sum_{n = 1}^{N} \log \left( \frac{\Q{(n+1)}}{\Q{(n)}} \right)\\
&\quad = \dsum{x}{y} \p{} \Q{*} \log \left(\frac{\Q{(N+1)}}{\Q{(1)}} \right)\\
& \quad \stackrel{(b)}\le \dsum{x}{y} \p{}  \Q{*} \log \left( \frac{\Q{*}}{\Q{(1)}} \right)
\end{align*}
where $(b)$ follows using the logarithm inequality.\\
Since at any iteration $n$, $W[\q{(n)}] - V[\q{*}] \ge 0$ and for all integers $N > 0$ the partial sum is upper-bounded by a constant $L(\q{*}, \q{(0)})$ dependent only on the initial probability assignment $\q{(0)}$, we have that $\lim_{N \to \infty} \sum_{n = 1}^{N} (W[\q{(n)}] - V[\q{*}])$ exists and it is finite hence $\lim_{n \to \infty} W[\q{(n)}] - V[\q{*}] = 0$. This completes the proof.

\section{Proof of Lemma \ref{lemma: Newton-Jacobian}} \label{proof: Newton-Jacobian}

    We first derive the functional form of $J_T$:
    \begin{align*}
        &\frac{\partial T[\q{(n)}, u](i)}{\partial u(j)} \\
        &= \delta_{i,j} - \q{(n)}(i) \ssum{x} \p{}(x) \frac{\frac{\partial \bA[i]{}}{\partial u(j)}}{\ssumh{k} \q{(n)}(k) \bA[k]{} } \\
        & \quad + \q{(n)}(i) \ssum{x} \p{}(x)\frac{\bA[i]{} \left(\ssum{k} \q{}(k) \frac{\partial \bA[k]{}}{\partial u(j)} \right)}{\left( \ssumh{k} \q{(n)}(k) \bA[k]{} \right)^2}\\
        &\frac{\partial \bA[i]{}}{\partial u(j)} = \bA[i]{}\left( -s_P \frac{\partial^2 D(\p{}||v)}{\partial v(i)^2}\Bigg{|}_u \right) \delta_{i,j}.
    \end{align*}
By defining matrices $M$, $\Gamma$ and $C$, respectively, as in \eqref{matrix: Newton_M}, \eqref{matrix: Newton_Gamma}, \eqref{matrix: Newton_C}, the matrix $J_T(\cdot)$ can be rewritten as in \eqref{eq: NewtonFunctionJacobian}.
To prove its invertibility, we need to ensure that $0$ is not part of the set of eigenvalues of $J_T(y)$, i.e., $ 0 \notin \eig(J_T(y)), ~ \forall y \in \mathbb{R}^{|\setSym|}$. 
Noticing that $M[\q{(n)}, u](i,j) \ge 0, \forall (i,j) \in \setSym^2$ and $\ssum{i} M[\q{(n)}, u](i,j) = C[\q{(n)}, u](i) $, we can define the sets $D_i$ as:
    \begin{align*}
        D_i = \Big\{ \lambda \in \mathbb{R}: &\left|\lambda - \left(C[\q{(n)},u](i) - M[\q{(n)}, u](i,i) \right)\right| \nonumber \\
        & \le C[\q{(n)},u](i) - M[\q{(n)}, u](i,i)\Big\}
    \end{align*} 
    Since $\bigcup_{i \in \setSym} D_i \subseteq \mathbb{R}_0^+$, we can apply Gershgorin Circle Theorem \cite{horn2013matrix} to prove that $C[\q{(n)},u] - M[\q{(n)}, u]$ has only non-negative eigenvalues.
    Moreover, since $D(\p{}||\cdot)$ is a convex function in its second argument, $\Gamma[\q{(n)}, u]$ is a positive semi-definite matrix. Therefore, we obtain
    \begin{align*}
        J_T[\q{(n)}, u] &= I + \left(C[\q{(n)}, u] - M[\q{(n)}, u]\right) \cdot \Gamma[\q{(n)}, u] \\
        &\ge I > 0
    \end{align*}
    This concludes the proof.

\section{Proof of Theorem \ref{th:RAM}} \label{proof:BA Convergence Approximate}

Let $V[\cdot]$, $W[\cdot]$ be the functionals defined in \eqref{eq: V} and \eqref{eq: W}, respectively. Moreover, let $\tW[\cdot]$ be a functional obtained by the alternating sequence  $ \tq{(n)} \to \tQ{(n+1)} \to \tq{(n+1)}$ substituting $\tQ{}$ with fixed $\tq{}$ as follows
\begin{align}
    \tW[\tq{(n)}] =& - \ssum{x} \p{}(x) \log \left( \ssumh{y} \tq{(n)}(y)\A{(n)} \right) \nonumber \\
    &+s_P \ssumh{y} \tq{(n+1)}(y) \frac{ \p{}(y) }{\va{(n)}(y)} \df \left(\frac{ \p{}(y) }{\va{(n)}(y)} \right) \nonumber\\
    &+ s_P \Bigg{[} \ssumh{y} \tq{(n+1)} f \left(\frac{ \p{}(y) }{\q{(n+1)}(y)} \right)  \nonumber\\
    &\qquad - \ssumh{y} \tq{(n+1)} f \left(\frac{ \p{}(y) }{\va{(n)}(y)} \right) \Bigg{]} \label{eq: W_tilde}
\end{align}
where $A^{(n)} = A[\va{(n)}]$. Similarly to Theorem \ref{th:OAM}, we let $\q{*}$ be any probability vector and $\Q{*} = \Q{}[\q{*}]$ such that $  R(D,P) = V[\q{*}] - s_DD - s_PP$ and consider that
\begin{align*}
    \begin{split}
        &\dsum{x}{y} \p{} \Q{*} \log \left( \frac{\tQ{(n)}}{\tQ{(n+1)}} \right) \\
        & \quad= \dsum{x}{y} \p{} \Q{*} \log \left( \frac{\tQ{(n)}}{\tq{(n)}} \right) \\
        &\qquad - \dsum{x}{y} \p{} \Q{*} \log \left( \A{(n)} \right)\\
        & \qquad + \dsum{x}{y} \p{} \Q{*} \log \left( \ssum{i} \tq{(n)}(i) \A[i]{(n)} \right).
    \end{split}
\end{align*}
Substituting the definition of $\A[]{}$ and using \eqref{eq: utility_derivation_1}, we obtain
\begin{align}
    \begin{split}
        &\dsum{x}{y} \p{} \Q{*} \log \left( \frac{\tQ{(n)}}{\tQ{(n+1)}} \right)  \le \quad V[\q{*}] - \tW[\tq{(n)}]\\
        & \quad  +s_P \Bigg[ \ssumh{y} \q{*}(y) f \left( \frac{ \p{}(y) }{\va{(n)}(y)} \right) - D_f(\p{}||\q{*}) \\
        &\qquad ~~~~~~ +\ssumh{y} \q{(n+1)}(y) \frac{ \p{}(y) }{\va{(n)}(y)} \df \left(\frac{ \p{}(y) }{\va{(n)}(y)} \right)\\
        &\qquad ~~~~~~ -\ssumh{y} \q{*}(y) \frac{ \p{}(y) }{\va{(n)}(y)} \df \left(\frac{ \p{}(y) }{\va{(n)}(y)} \right)  \\
        & \qquad ~~~~~~ + \ssumh{y} \tq{(n+1)} f \left(\frac{ \p{}(y) }{\q{(n+1)}(y)} \right)  \\
        & \qquad ~~~~~~ -\ssumh{y} \tq{(n+1)} f \left(\frac{ \p{}(y) }{\va{(n)}(y)} \right) \Bigg].\label{rhr_eq}
    \end{split}
\end{align}
Note that the right-hand side of \eqref{rhr_eq} can be bounded by
{ \allowdisplaybreaks
\begin{align*}
    \begin{split}
        &\ssumh{y} \q{*}(y) f\left(\frac{ \p{}(y) }{\va{(n)}(y)}\right) - D_f \left(\p{}||\q{*} \right) \\
        & \quad + \ssumh{y} \q{(n+1)}(y) \frac{ \p{}(y) }{\va{(n)}(y)} \df \left(\frac{ \p{}(y) }{\va{(n)}(y)} \right) \\
        & \quad - \ssumh{y} \q{*}(y) \frac{ \p{}(y) }{\va{(n)}(y)} \df \left(\frac{ \p{}(y) }{\va{(n)}(y)} \right) \\
        & \quad + \ssumh{y} \tq{(n+1)} f \left(\frac{ \p{}(y) }{\q{(n+1)}(y)} \right) - \ssumh{y} \tq{(n+1)} f \left(\frac{ \p{}(y) }{\va{(n)}(y)} \right)\\
        & = D_f \left(\p{}||\va{(n)} \right) - D_f \left( \p{}||\q{*} \right) \\
        & \quad + \ssum{i} (\q{*} - \va{(n)}) \left[ f \left(\frac{ \p{}(y) }{\va{(n)}(y)} \right) - \frac{ \p{}(y) }{\va{(n)}(y)}\df \left(\frac{ \p{}(y) }{\va{(n)}(y)} \right) \right] \\
        & \quad + D_f \left(\p{}||\q{(n+1)} \right) - D_f \left(\p{}||\va{(n)} \right) \\
        &\quad + \ssum{i} \left(\va{(n)} - \q{(n+1)} \right) f \left(\frac{ \p{}(y)}{\va{(n)}(y)}\right)\\
        &\quad - \ssum{i} \left(\va{(n)} - \q{(n+1)} \right) \frac{ \p{}(y) }{\va{(n)}(y)}\df \left(\frac{ \p{}(y)}{\va{(n)}(y)}\right)\\
        & = - \left[ D_f \left(\p{}||\q{*}\right) - T_{D_f(p||\cdot),\va{(n)}}(\q{*})\right] \\
        & \quad + \left[D_f(\p{}||\q{(n+1)}) - T_{D_f(p||\cdot),\va{(n)}}\left(\q{(n+1)} \right)\right] \\
        & \stackrel{(a)}\le D_f\left(\p{}||\q{(n+1)}\right) - T_{D_f(p||\cdot),\va{(n)}}\left(\q{(n+1)} \right)
    \end{split}
\end{align*}
}where $T_{D_f(p||\cdot),\va{(n)}}\left(\q{(n+1)} \right)$ is the first order Taylor expansion of $D_f(\p{}||\cdot)$, centered in $\va{(n)}$ and evaluated in $\q{(n+1)}$,  and $(a)$ is verified since $D_f(\p{}||\cdot)$ is a convex function in its second argument.
\par Since $\tW[\tq{(n)}] \ge W[\tq{(n)}] \ge V[\q{*}]$ we can rewrite
\begin{align*}
    \begin{split}        
        0 \le \tW[\tq{(n)}] - V[\q{*}] \le G[\tq{(n)}]
    \end{split}
\end{align*}
where
\begin{align*}
 \begin{split}        
        & G[\tq{(n)}] = \dsum{x}{y} \p{} \Q{*} \log \left( \frac{\tQ{(n+1)}}{\tQ{(n)}} \right)  \\
        & \qquad + s_P \left[ D_f \left( \p{}||\q{(n+1)} \right) - T_{D_f(p||\cdot),\va{(n)}}\left(\q{(n+1)} \right) \right].
    \end{split}
\end{align*}
Summing over $N$ terms we obtain
\begin{align*}
\begin{split}
        &\sum_{n = 1}^{N}\tW[\tq{(n)}] - V[\q{*}] \\
        & \le  \dsum{x}{y} \p{} \Q{*} \sum_{n = 1}^{N} \log \left(\frac{\tQ{(n+1)}}{\tQ{(n)}} \right)\\
        & \quad  + s_P \sum_{n = 1}^{N} D_f \left(\p{}||\q{(n+1)} \right) - T_{D_f(p||\cdot),\va{(n)}}\left(\q{(n+1)}\right)\\
        & \le \dsum{x}{y} \p{}  \Q{*} \log \left(\frac{\Q{*}}{\tQ{(1)}}\right) + \sum_{n = 1}^{N} o \left(||\q{(n+1)} - \va{(n)}|| \right)\\
        & \le \tilde{L}(\q{*}, \q{(0)}) 
    \end{split}
\end{align*}
where $\tilde{L}(\q{*}, \q{(0)})$ is finite if the limit $\lim_{n \to \infty} ||\q{(n+1)} - \va{(n)}|| = 0$ converges at least linearly. Thus, we can rewrite
\begin{align*}
            0 \le \lim_{N \to \infty}\sum_{n = 1}^{N} \tW[\tq{(n)}] - V[\q{*}] \le \tilde{L}(\q{*}, \q{(0)})
\end{align*}
proving the convergence  as in Theorem \ref{th:OAM}. This completes the proof.

\section{Proof of Theorem \ref{th: UpperAndLowerboundToRate}} \label{proof: UpperAndLowerboundToRate}

We start by first introducing the following auxiliary lemma.

\begin{lemma} \label{lemma: RDP lowerbound}
 Let $s_D \ge 0$, $s_P \in [0, s_{P, \max})$ be given with $s=(s_D,s_P)$ and let $\Q{}$ be a transition matrix included in the set $\mathcal{L}_{(D,P)}$ defined as follows:
\begin{align*} 
    \mathcal{L}_{(D,P)} = \{ \Q{}: \E[\Q{}]{d(X,Y)} \le D \land D_f(\p{}|| \Qm{}) \le P \}.
\end{align*}
where $q_Y = \ssum{x} \p{}\Q{}$. Furthermore, let $\Lambda_{s,\va{}[\Qm{}]}$ be the set defined as
\begin{align*}
    \Lambda_{s,\va{}[\Qm{}]}\triangleq \Big\{ \lambda \in \realR^{|\setSym|}:& \forall x \in \setSym, \lambda(x) \ge 0  \land \forall y \in \setSym, \\
    & \ssum{x} \p{}(x) \lambda(x) A[v[q_Y]](x,y,s) \le 1 \Big\}.
\end{align*}
Then, $\forall \mathbf{\lambda} \in \Lambda_{s,\va{}[\Qm{}]}$, we obtain
\begin{align*} 
\begin{split}
    R(D,P) \ge & \ssum{x} \p{}(x) \log\left(\lambda(x)\right) - s_D D \\
    &-s_P\dsum{x}{y} \p{}(x) \Q{}(x,y) g(\p{}(y), v[q_Y](y)).
\end{split}
\end{align*}
\end{lemma}

\begin{IEEEproof} 
    Let $\mathbf{\lambda} \in \Lambda_{s,\va{}[\Qm{}]}$ and $\Q{} \in \mathcal{L}_{(D,P)}$, then:
    \begin{align*}
        \begin{split}
                &I\left(\p{},\Q{} \right)  + \ssum{x} \p{}(x) \log\left(\frac{1}{\lambda(x)}\right) + s_D D \\
                &\quad + s_P \dsum{x}{y} \p{}(x) \Q{}(x,y) g(\p{}(y), v[q_Y](y))\\
                &\ge \dsum{x}{y} \p{} \Q{} \log \left(\frac{\Q{}}{\Qm{} \lambda(x)A[v[q_Y]](x,y,s) } \right)\\
                &\ge 1 - \dsum{x}{y} \p{} \Q{} \left(\frac{\Qm{}\lambda(x)A[v[q_Y]](x,y,s)}{\Q{}}\right)\\
                & = 1 - \ssumh{y} \Qm{} \ssum{x} \p{} \lambda(x) A[v[q_Y]](x,y,s)\\
                &\ge 1 - \ssumh{y} \Qm{} = 0.
        \end{split}
    \end{align*}
    The equality is ensured by the possibility of choosing $\lambda(x) \in \Lambda_{s,\va{}[\Qm{}]}$ as
    \begin{align*}
        \begin{split}
            \lambda(x) = \frac{1}{\ssumh{y} q_Y(y) A[v[q_Y]](x,y,s)}\\
        \end{split}
    \end{align*}
    that once substituted describes the optimization problem found in Corollary \ref{cor: Minimization Conditions}. This completes the proof of the lemma.
\end{IEEEproof}

Now we use Lemma \ref{lemma: RDP lowerbound} to prove Theorem \ref{th: UpperAndLowerboundToRate}. In particular, \eqref{eq: Upperbound Approximation} can be derived from the following inequality:
\begin{align*}
    \begin{split}
        &R(D,P)  \le \quad I \left(\p{},\tQ{}[\tq{}] \right)\\
        & = \dsum{x}{y} \p{}(x) \tQ{}[\tq{}](y|x)  \Bigg[ \log \left(\tQ{}[\tq{}](y,x)\right) \\
        & ~~~~~~~~~~~~~~~~~~~~~~~~~~~~-\log\left({\ssum{x} \p{}(x) \tQ{}[\tq{}](y|x)} \right) \Bigg]\\
        & = -s_D D - s_P P - \ssumh{y} \tq{}(y) \cx[] \log(\cx) \\
        & \quad + s_P \ssumh{y} \Qm{}[\tq{}](y) \frac{ \p{}(y) }{\va{}[\tq{}](y)} \df \left(\frac{ \p{}(y) }{\va{}[\tq{}](y)} \right)\\
        & \quad - s_P \Bigg\{\ssumh{y} \Qm{}[\tq{}](y) f \left( \frac{ \p{}(y)}{\va{}[\tq{}](y)} \right) - P \Bigg\} \\
        &\quad - \ssum{x} \p{}(x) \log \left( \ssumh{y} \tq{}(y)A[v[\tq{}]](x,y,s) \right) \\
        & = -s_D D - s_P P - \ssumh{y} \tq{}(y) \cx[] \log(\cx) + \tW[\tq{}].
    \end{split}
\end{align*}
(\ref{eq: Lowerbound Approximation}) is derived as an application of Lemma \ref{lemma: RDP lowerbound} by choosing $\lambda(x)$ as 
\begin{align*}
    \lambda(x) = \left( c_{\max} \ssumh{y} \tq{}(y) A[v[\tq{}]](x,y,s) \right)^{-1}
\end{align*}
which respects the assumption of the theorem. This completes the proof.

\section{Proof of Theorem \ref{th: base_alg_Jacobian}} \label{proof: base_alg_Jacobian}

The functional form of $J[\q{}](i,j)$ in the case of Theorem \ref{th:OAM} is obtained from
\begin{align*} 
    \begin{split}
        &\frac{\partial S[\q{}](i)}{\partial \q{}(j)} = c[\q{},S[\q{}]](i) \delta_{i,j} \\
        &\quad + \q{}(i) \ssum{x} \p{}(x) \frac{\partial}{\partial \q{}(j)} \left( \frac{A[S[\q{}]](x,i,s)}{\ssum{k} \q{}(k) A[S[\q{}]](x,k,s)} \right) 
    \end{split}
\end{align*}
where
\begin{align*} 
    \begin{split}
        &\frac{\partial (\ssum{k} \q{}(k) A[S[\q{}]](x,k,s)) }{\partial \q{}(j)} = \\ 
        & \qquad   \ssumh{k} \q{}(k) \frac{\partial A[S[\q{}]](x,k,s)}{\partial \q{}(j)} + A[S[\q{}]](x,j,s)
    \end{split}\\
    \begin{split}
        &\frac{\partial \A[i]{}}{\partial \q{}(j)} = \\
        & \qquad -s_P A[S[\q{}]](x,i,s) \frac{\p{}(i)^2}{(S[\q{}(i)])^3} \df \left( \frac{\p{}(i)}{S[\q{}(i)]} \right) \frac{\partial S[\q{}](i)}{\partial \q{}(j)}
    \end{split}.
\end{align*}
By noticing that
\begin{align*}
 \frac{\p{}(i)^2}{(S[\q{}](i))^3} \df \left( \frac{\p{}(k)}{S[\q{}](i)} \right) = \frac{\partial^2}{\partial q(i)^2} D_f(\p{}||q)\Big|_{S[\q{}]}  
\end{align*}
 and defining the matrices $M$ and $\Gamma$ as in \eqref{matrix: Newton_M} and \eqref{matrix: Newton_Gamma}, we can rewrite the entries of $J[\q{*}]$ as
\begin{align} 
        J[\q{*}](i,j) &= c[\q{*}, \q{*}](i) \left( \delta_{i,j} - \Gamma[\q{*}, \q{*}](i) J[\q{*}](i,j) \right) \nonumber\\
        & \quad +\ssumh{k} \Gamma[\q{*}, \q{*}](k) M[\q{*},\q{*}](i,k) J[\q{*}](k,j) \nonumber \\
        &\quad - M[\q{*},\q{*}](i,j) \label{eq: Jacobian Entries}
\end{align}
where $\delta_{i,j}$ is the Kronecker delta. As a result, we can express \eqref{eq: Jacobian Entries} in matrix form as follows
\begin{align} 
    \begin{split}
        J[\q{*}] &= C[\q{*}, \q{*}]( I - \Gamma[\q{*}, \q{*}] J[\q{*}]) \\
        & \quad - M[\q{*},\q{*}] + M[\q{*},\q{*}] \Gamma[\q{*},\q{*}] J[\q{*}] \nonumber\\
        &= (C[\q{*}, \q{*}] - M[\q{*},\q{*}])(I - \Gamma J[\q{*}]) \label{eq: Jacobian_tmp1}
    \end{split}
\end{align}
where $C[\cdot,\cdot]$ is defined in \eqref{matrix: Newton_C}. Finally, we obtain \eqref{eq: JacobianEq1} noticing that $C[\q{*},\q{*}] = I$ due to the optimality conditions found in Lemma \ref{lemma: opt_condition_q}, thus concluding the proof.

\section{Proof of Lemma \ref{lemma: lowerboundEigM}} \label{proof: loweboundEigM}
Let matrices $\Phi$ and $Q$ be defined as
\begin{align*}
    \Phi &\triangleq \left[ \sqrt{\p{}(x)} \frac{ A[\q{*}](x,i,s)}{\ssum{k} \q{}(k) A[\q{*}](x,k,s)}  \right]_{(i,x) \in \setSym^2} \\
    Q &\triangleq \diag \Big[\q{*}(i)\Big]_{i \in \setSym}.
\end{align*} 
Then, the following identity can be verified
\begin{align*}
    Q^{\frac{1}{2}} M^* Q^{-\frac{1}{2}} = Q^{\frac{1}{2}} \Phi \Phi^T Q^{\frac{1}{2}} = ( Q^{\frac{1}{2}} \Phi)( Q^{\frac{1}{2}} \Phi)^T
\end{align*}
where $Q^{\frac{1}{2}} M^* Q^{-\frac{1}{2}}$ is necessarily symmetric and at least semi-positive definite. 
To guarantee positive definiteness of $Q^{\frac{1}{2}} M^* Q^{-\frac{1}{2}}$, and thus the fact that the eigenvalues of $M$ are strictly positive, we need to impose conditions on the full rank of $\Phi$. To address them, we can factorize $\Phi$  into the product $ \Phi = U D V$, where
\begin{align*}
    D &= \left[ e^{ - s_D d(i,j)} \right]_{(i,j) \in \setSym \times \setSym}\\
    U & = \diag \left[e^{- s_P g(\p{}, \q{*},i)} \right]_{i \in \setSym}\\
    V & = \diag \left[\frac{\sqrt{\p{}(x)}}{\ssum{k} \q{}(k) A[\q{*}](x,k,s)} \right]_{x \in \setSym}.
\end{align*}
Since both $U$ and $V$ are positive definite matrices, it is easy to verify that $\Phi$ is a full-rank matrix if and only if $D$ is full rank too. This completes the proof.

\section{Proof of Lemma \ref{lemma: upperboundEigM}} \label{proof: upperboundEigM}
We can verify that the lemma is the result of the Gershgorin Circle Theorem \cite{horn2013matrix} applied to the columns of $M^*$. Noticing that all entries $M^*(i,j)$ are strictly positive, the disk radius $R(j)$ for column $j$ is:
\begin{gather*}
    R(j) + M^*(i,i) = \ssumh{i} M^*(i,j)
\end{gather*}
where
\begin{align*}
    &\ssumh{i} M^*(i,j)  \\
    &= \ssumh{i} \q{*}(i) \ssum{x} \p{}(x) \frac{A[\q{*}](x,i,s) A[\q{*}](x,j,s){}}{(\ssum{k} \q{*}(k) A[\q{*}](x,k,s))^2}\\
    &= \ssum{x} \p{}(x) \frac{A[\q{*}](x,j,s) \left(\ssumh{i}  \q{*}(i) A[\q{*}](x,i,s) \right)}{ \left(\ssum{k} \q{*}(k) A[\q{*}](x,k,s)\right)^2} \\
    &= c[\q{*}, \q{*}](j) = 1.
\end{align*}
Thus the eigenvalues of $M^*$ are each in at least one of the disks $I_{i} = \{ z \in \realR: |z - M^*(i,i)|\le 1 - M^*(i,i)\}$, which are all contained in the disk $I = \{ z \in \realR: |z|\le 1\}$. This completes the proof.

\section{Proof of Theorem \ref{th: EXPConvergenceBase}} \label{proof: EXPConvergenceBase}
Using Lemmas \ref{lemma: lowerboundEigM} and \ref{lemma: upperboundEigM}, the following inequalities hold 
\begin{align*}
    0 < \eig(M^*) \le 1 &\implies 0 \le \eig( I - M^*) < 1\\
    0 \le \eig(I - M^*) < 1 &\xRightarrow{(a)} 1 \le \eig(I + (I-M^*)\Gamma)\\
    1 \le \eig(I + (I-M^*)\Gamma) &\implies\\
    &0 < \eig((I + (I-M^*)\Gamma^*)^{-1}) \le 1
\end{align*}
where $(a)$ is due to $\Gamma$ being a positive definite matrix. Using the previous inequalities, we can rewrite \eqref{eq: JacobianEq1} as follows:
\begin{align*}
    J(\q{*}) = (I + (I - M^*)\Gamma^*)^{-1}(I - M^*).
\end{align*}
Define $\theta_{\sup} \triangleq \eig_{\max}(I - M^*) \cdot \eig_{\max}((I + (I - M^*)\Gamma^*)^{-1}) $. Then, we can show that $0  \le \eig(J[\q{*}]) \le \theta_{\sup} < 1$ is always verified. 
The second part of the theorem follows directly from \cite[Theorem 5]{9476038} hence we omit it. This completes the proof.

\section{Proof of Theorem \ref{th: approximation_exponential_interval}} \label{proof: approximation_exponential_interval}

Since $R(D,P)$ is a non-increasing convex function in $D_s$ and $P_s$, we can derive:
\begin{align*}
    \frac{\partial^2 R(D,P) }{ \partial P_s^2} = - \frac{\partial s_P }{ \partial P_s} \ge 0
\end{align*}
meaning that more constrained values of $P_s$ are associated with larger $s_P$. Thus, for a given $s_D$, let $s_{P,\max}$ be the value of the Lagrangian $s_P$ associated with the constraint $P = 0$. Then the solution $\q{*}$ is necessarily unique and must be $\q{*} = \p{}$. Then, due to the properties of the Jacobian $J(\q{*})$,
\begin{align*}
    \begin{split}
        (I - \Gamma J(\q{*})) \ge 0  \implies s_{P,\max} \le \frac{1}{\theta_{\max} f''(0)}.
    \end{split}
\end{align*}
In order to guarantee $0 \le \{\theta_{a,i}\}_{i \in \setSym} < 1$, it is sufficient to have
\begin{align*}
    \begin{split}
        (I - \Gamma) \ge 0  \implies s_{P,\max} \le \frac{1}{f''(0)}.
    \end{split}
\end{align*}
Since in the non-degenerate case we have $0 < \theta_{\max} < 1$, we can construct $\epsilon' = \min \left\{s_{P,\max},\frac{1}{f''(0)} \left(\frac{1}{\theta_{\max}} - 1 \right) \right\}$ > 0 and define the set $I_{s_P}^{\epsilon'}$ where the exponential convergence of the approximate algorithm is guaranteed. This concludes the proof.

\section{Proof of Lemma \ref{lemma: LowboundTV}} \label{proof: LowboundTV}
We remind that $f = |x - 1|$ is the function associated with the TV distance $TV(\cdot||\cdot)$. The first statement can be proved by first establishing the uniform convergence of $f_n \to f$ as $n\to \infty$
    \begin{align*}
        &\sup_{x \in \realR} \left|f_n(x) - f(x)\right| \\
        &= \sup_{x \in \realR} \left| \frac{2}{\pi}(x-1) \arctan(n(x-1)) - |x - 1| \right|\\
        &= \sup_{x \in \realR^+_{0}} \left| \frac{2}{\pi}x \arctan(nx) - x\right| \\
        &= \sup_{x \in \realR^+_{0}} \frac{2}{\pi}x \left(\frac{\pi}{2} - \arctan(nx)\right)\\
        &= \sup_{x \in \realR^+_{0}} \frac{2}{\pi}x\arctan\left(\frac{1}{nx}\right) \le \sup_{x \in \realR^+_{0}} \frac{2}{n\pi} = \frac{2}{n\pi}
    \end{align*}
    meaning that $\lim_{n \to \infty} \sup_{x \in \realR} |f_n(x) - f(x)| = 0$. A direct consequence of the above is that  $D_{f_n} \to TV$ uniformly in the limit of $n \to \infty$ since, for any $p,q \in \mathcal{P}(\setSym)$, 
    \begin{align*}
        &\lim_{n \to \infty} |D_{f_n}(p||q) - TV(p||q)| \\ &=  \lim_{n \to \infty} \left|\ssum{x} q(x) \left(f_n\left(\frac{p(x)}{q(x)}\right) - f\left(\frac{p(x)}{q(x)}\right) \right)\right|\\
        & \le  \ssum{x} q(x) \lim_{n \to \infty} \left|f_n\left(\frac{p(x)}{q(x)}\right) - f\left(\frac{p(x)}{q(x)}\right)\right| = 0.
    \end{align*}
Instead, the inequality $f_n(x) \le f(x), \forall x \in \realR$, implies that for all $n \in \mathbb{N}$ and $\forall p,q \in \mathcal{P}(\setSym)$, the inequality $D_{f_n}(p||q) \le TV(p||q)$  holds. This concludes the proof.

\bibliographystyle{IEEEtran}
\bibliography{string,biblio}

\begin{IEEEbiographynophoto}
{Giuseppe Serra} (S'22) received his B.Sc degree in electronic and communication engineering from Politecnico di Torino (PoliTo), Italy, in 2019, and his double M.Sc. degrees in telecommunication engineering from Politecnico di Torino (PoliTo), Italy, and data science from EURECOM, France, in 2022. He is currently a PhD student at the Communication Systems Department, EURECOM, France. His research focuses on information theory and optimization, with a particular interest in the topic of goal-oriented semantic communication.
\end{IEEEbiographynophoto}

\begin{IEEEbiographynophoto}
{Photios A. Stavrou} (S'10-M'16-SM'22) received his D. Eng in 2008 from the Department of Electrical and Computer Engineering (ECE) of the Faculty of Engineering at Aristotle University of Thessaloniki, Greece, and his Ph.D. degree in 2016 from the Department of ECE of the Faculty of Engineering at University of Cyprus, Cyprus. From November 2016 to September 2022, he held post-doctoral positions at Aalborg University in Denmark, at KTH Royal Institute of Technology in Sweden, and at EURECOM in France. As of September 2022, Dr Stavrou is an Assistant Professor in the Communication Systems Department at EURECOM, Sophia-Antipolis, France. Dr Stavrou and his co-authors are the recipients of the best paper award in industry innovation in ACP/POEM 2023. His research interests span information and communication theories, networked control systems theory, goal-oriented communication, feedback and privacy in communication, optimization, and game theory.
\end{IEEEbiographynophoto}

\begin{IEEEbiographynophoto}{Marios Kountouris} (S’04–M’08–SM’15–F’23) received the diploma degree in electrical and computer engineering from the National Technical University of Athens (NTUA), Greece in 2002 and the M.S. and Ph.D. degrees in electrical engineering from Télécom Paris, France, in 2004 and 2008, respectively. He is currently a Professor in the Communication Systems Department at EURECOM, France, and a Distinguished Researcher (Research Professor) in the Department of Computer Science and Artificial Intelligence and the Andalusian Research Institute in Data Science and Computational Intelligence (DaSCI) at the University of Granada, Spain. Prior to his current appointments, he held positions at CentraleSupélec, France, the University of Texas at Austin, USA, the Huawei Paris Research Center, France, and Yonsei University, South Korea. He is the recipient of a Consolidator Grant from the European Research Council (ERC) in 2020 on goal-oriented semantic communication. He has served as an Editor for the IEEE Transactions on Wireless Communications, the IEEE Transactions on Signal Processing, and the IEEE Wireless Communication Letters. He has received several awards and distinctions, including the 2022 Blondel Medal, the 2020 IEEE ComSoc Young Author Best Paper Award, the 2016 IEEE ComSoc CTTC Early Achievement Award, the 2013 IEEE ComSoc Outstanding Young Researcher Award for the EMEA Region, the 2012 IEEE SPS Signal Processing Magazine Award, the IEEE SPAWC 2013 Best Paper Award and the IEEE Globecom 2009 Communication Theory Best Paper Award. He is a Fellow of the IEEE, the AAIA, and the AIIA Fellow, and a Professional Engineer registered with the Technical Chamber of Greece.
\end{IEEEbiographynophoto}

\end{document}